\documentclass[12pt,draftclsnofoot,onecolumn]{IEEEtran}

\IEEEoverridecommandlockouts
\usepackage{indentfirst,flushend}
\usepackage{amssymb,bm,mathrsfs,times,amscd,amsmath,amsthm,amsfonts,bbm}
\usepackage{latexsym}
\usepackage{dsfont,slashbox}
\usepackage{graphicx}
\usepackage{subfigure}
\usepackage{color}
\usepackage{cases}

\newtheorem{remark}{Remark}
\usepackage{multirow}
\usepackage[sort]{cite}
\usepackage{multicol}
\usepackage[nooneline,flushleft]{caption2}
\usepackage{clrscode}
\usepackage{algorithm}
\usepackage{psfrag,stfloats,nicefrac}
\usepackage{amsmath}
\usepackage{amssymb}

\usepackage{algorithmic}
\usepackage{supertabular}
\usepackage{makecell}
\usepackage{epstopdf}
\usepackage{threeparttable}
\usepackage{ulem}  
\allowdisplaybreaks[4]

\begin{document}

\title{GBLinks: GNN-Based Beam Selection and Link Activation for Ultra-dense D2D mmWave Networks}
\author{Shiwen~He,~\IEEEmembership{Member,~IEEE}, Shaowen Xiong, Wei Zhang,~Yiting Yang,~Ju Ren,~\IEEEmembership{Member,~IEEE}, and Yongming~Huang,~\IEEEmembership{Senior Member,~IEEE} 
\thanks{S. He, S. Xiong, W. Zhang, and J. Ren are with the School of Computer Science and Engineering, Central South University, Changsha 410083, China. S. He is also with the National Mobile Communications Research Laboratory, Southeast University, and the Purple Mountain Laboratories, Nanjing 210096, China. (email: \{shiwen.he.hn, shaowen.xiong, renju\}@csu.edu.cn, sunbirdcsu@Outlook.com).}
\thanks{Y. Yang is also with the Purple Mountain Laboratories, Nanjing 210096, China.  (email: yangyiting@pmlabs.com.cn).}
\thanks{Y. Huang is with the National Mobile Communications Research Laboratory, School of Information Science and Engineering, Southeast University, Nanjing 210096, China. He is also with the Purple Mountain Laboratories, Nanjing 210096, China. (email: huangym@seu.edu.cn). }
}

\maketitle
\vspace{-.6 in}

\begin{abstract}
In this paper, we consider the problem of joint beam selection and link activation across a set of communication pairs to effectively control the interference between communication pairs via inactivating part communication pairs in ultra-dense device-to-device~(D2D) mmWave communication networks. The resulting optimization problem is formulated as an integer programming problem that is nonconvex and NP-hard. Consequently, the global optimal solution, even the local optimal solution, cannot be generally obtained. To overcome this challenge, this paper resorts to design a deep learning architecture based on graph neural network to finish the joint beam selection and link activation, with taking the network topology information into account. Meanwhile, we present an unsupervised Lagrangian dual learning framework to train the parameters of the GBLinks model. Numerical results show that the proposed GBLinks model can converges to a stable point with the number of iterations increases, in terms of the weighted sum rate. Furthermore, the GBLinks model can reach near-optimal solution through comparing with the exhaustive search scheme in small-scale ultra-dense D2D mmWave communication networks and outperforms GreedyNoSched and the SCA-based method. It also shows that the GBLinks model can generalize to varying densities and coverage regions of ultra-dense D2D mmWave communication networks.
\end{abstract}


\section*{\sc \uppercase\expandafter{\romannumeral1}. Introduction}


Millimeter wave~(mmWave) communication is a key enable technology for future wireless networks, which can address the challenge of spectrum shortage. However, mmWave signals encounter serious pathloss due to the large rain attenuation and Oxygen attenuation, etc. In order to make up for this shortcoming, large-scale antenna array providing sufficient antenna gain is adopted in mmWave communication system~\cite{ProcShi2021}. However, for mmWave communication system, the conventional digital beamforming techniques are not suitable because they require each antenna element to have dedicated radio frequency (RF) link, which is expensive and consumes too much energy. Hybrid beamforming, consisting of an analog and a digital beamforming, is a cost-effective alternative, which can significantly reduce the hardware cost and power consumption by using a small number of RF links~\cite{OJCSZhang2020}.

Recently, the researches on the design of hybrid beamforming for mmWave communication have attracted extensive attentions in both academia and industry.
O. E. Ayach $et~al.$ exploited the sparsity of mmWave channels to investigate the design of hybrid precoder for maximizing the throughput of a point-to-point communication system~\cite{el2014spatially}.  X. Gao $et~al.$ investigated the energy-efficient design of large-scale antenna array mmWave communication systems~\cite{gao2016energy}. S. He $et~al.$ studied the design of hybrid precoder for the downlink cache-enabled mmWave communication networks~\cite{8633840}. L. Zhao $et~al.$ investigated an energy efficient hybrid beamforming architecture with low-resolution A/Ds equipped at both the transmitter and the receiver for multi-user mmWave communication~\cite{9110851}. In addition, the directional transmission of mmWave communication helps to solve the serious interference problem and improves the system throughput of wireless networks~\cite{TWCVen2016}.

The fundamental problems are the user scheduling and the beamforming in multi-user multi-input multi-output (MIMO) wireless communication systems. P. C. Weeraddana $et~al.$ reviewed the weighted sum-rate maximization problem in many network control and optimization methods, such as power control, link scheduling, cross-layer control, and network utility maximization~\cite{sumrateMax2012}. Furthermore, generally speaking, the existing design method of beamforming is based on a given scheduled user set. Consequently, beamforming alone cannot generally achieve the optimal performance of ultra-dense mmWave communication networks. This implies that an effective joint beam design and user scheduling~(link activation) method is necessary for ultra-dense mmWave communication networks. J. Yu $et~al.$ investigated the problem of the maximum number of links that can be successfully scheduled simultaneously under Rayleigh-fading and multiuser interference without considering the problem of beam design.~\cite{9099107}. M. Ge $et~al.$ considered the multiuser MIMO scheduling problem with fixed beamforming vectors for dense wireless networks with access point cooperation~\cite{8422380}. Y. Niu $et~al.$ investigated the path planning and concurrent transmission algorithms for D2D mmWave communication network with fixed transmission beams~\cite{TCOMNiu2018}. Note that the aforementioned literature does not consider the problem of joint beam design and user scheduling for multi-antenna communication systems. Recently, S. He $et~al.$ considered the joint optimization of analog beam selection and user scheduling based on limited effective channel state information~(CSI) for a single cell downlink multi-user multiple-input single-output~(MISO) network~\cite{he2017joint}. The aforementioned references designed optimization algorithms to directly solve the original optimization problem or its variants. However, they do not exploit the information hidden in the wireless data to explore the transmission schemes for wireless communication networks.


To address the challenges faced by the traditional optimization methods, machine learning is becoming a powerful method to put intelligence into wireless networks, which can extract the patterns from wireless network topology and complex radio conditions. There are mainly two paradigms on this topic. The first one is ``end-to-end learning" directly employing a neural network to approximate the near-optimal solution of an optimization problem. C. Xu $et~al.$ investigated the analog beam selection at transmitter and user scheduling strategy based on multi-agent reinforcement learning for the downlink of multicell mmWave communication network~\cite{9104386}. J. Zhang $et~al.$ formulated the problem of beam alignment and tracking as a stochastic bandit problem and proposed two efficient algorithms to address the problem considered~\cite{Beam2020Zhang}. H. Sun $et~al.$ used a multi-layer perceptron~(MLP) to approximate the input-output mapping of the classical weighted minimum mean square error algorithm to speed up the computation~\cite{Sun2018Learning}. J. Tao $et~al.$ proposed a deep neural network-based hybrid beamforming algorithm for multi-user mmWave massive MIMO systems~\cite{8969154}. The second paradigm is ``learning and optimization", which uses neural networks instead of traditional algorithms to learn more difficult strategies. Machine learning technique was used to replace the pruning strategy in the branch-and-bound algorithm~\cite{shen2019lorm}. K. Lee $et~al.$ designed an iterative algorithm based on a typical optimization technique and proposed a learning algorithm based on a neural network to jointly optimize the transmit power and energy harvesting time to maximize the energy efficiency of the network~\cite{Lee2020Learning}.

In order to improve the performance and generalization ability of machine learning models, an effective idea is to incorporate the network topology information into the architecture of learning models avoiding learning the network topology from the data. W. Cui $et~al.$ showed that by using a deep learning approach, it is possible to bypass the channel estimation and to schedule links efficiently based solely on the geographic locations of communication pairs~\cite{JSACCui2019}. On the other hand, graph neural networks~(GNNs) have shown good performance in non-Euclidean scenarios in recent years, which can effectively exploit non-Euclidean data, e.g., CSI~\cite{Deep2020Zhang}. Y. Shen $et~al.$ utilized GNNs to develop scalable methods for solving the power control problem in $K$-user interference channels~\cite{shen2019graph}. They also identified a family of neural networks message passing GNNs (MPGNNs), and demonstrated that the radio resource management problems can be formulated as graph optimization problems enjoying a universal permutation equivalence property~\cite{shen2020graph}. They also took power control for $K$ single-antenna communication pairs and beamforming for $K$ communication pairs as two examples to analyze the performance and generalization of MPGNN-based methods. To solve the problem of link scheduling, M. Lee $et~al.$ constructed a fully-connected graph for the D2D network, and then proposed a novel graph embedding-based method to address the link scheduling problem without requiring the accurate CSI~\cite{ArxivLee2019}. Numerical results demonstrate that the proposed method is near-optimal compared with the existing state-of-art methods but is with only hundreds of training samples. M. Eisen $et~al.$ introduced the random edge graph neural network~(REGNN) in the wireless ad-hoc network, which performs convolutions over random graphs formed by the fading interference patterns in the wireless network~\cite{Eisen2020Optimal}. Numerical results demonstrate that REGNN is an effective parameterization for resource allocation policies for large-scale wireless networks.

Directional transmission has the ability to suppress interference, but the interference between communication pairs may still be serious in ultra-dense D2D mmWave communication networks, in which a large number of communication pairs want to simultaneously establish communication links on the same time-frequency resource. In this paper, we consider the problem of joint beam selection and link activation to effectively control the interference between communication pairs via inactivating part communication pairs in ultra-dense D2D mmWave communication networks. But, the authors of~\cite{he2017joint} and~\cite{9104386} only jointly consider the selection of analog beams at the transmitter and user scheduling for the downlink single-cell multi-user MISO network. When considering the problem of joint analog beam selection at communication pairs and user scheduling, it becomes more challenging for ultra-dense D2D mmWave communication networks. To overcome the difficulties encountered in solving the problem of joint beam selection and link activation, we formulate the problem of interest as a combinatorial optimization problem by introducing the indicator variables of beam selections for ultra-dense D2D mmWave communication networks. Then, a Lagrangian dual learning framework is proposed to train an end-to-end deep learning model designed based on GNN, called GBLinks, to solve the considered optimization problem. The main contributions are listed as follows
\begin{itemize}
  \item The joint beam selection and link activation problem is described as a constrained combinatorial optimization problem aiming at maximizing the total throughput. The variables needed to be optimized are the beam selection indicators of transmitter and receiver. Meanwhile, these variables are used to indirectly indicate the link activation in ultra-dense D2D mmWave communication networks.
  \item A Lagrangian dual learning framework~(LDLF) is proposed to train the GBLinks model in an unsupervised manner, which is utilized to parameterize the GBLinks model and satisfy the constraints we considered. Then the GBLinks model is designed based on GNNs to generate the beam selection and link activation policies, namely, the prediction of beam selection indicators.
  \item We also propose an optimization method based on successive convex approximation~(SCA) to address the formulated problem. We further discuss the scheme of the initialization and the update of Lagrangian multipliers of the provided algorithms.
  \item We use unlabeled data set to verify the effectiveness of LDLF, performance, and generalization ability evaluation of the GBLinks model. The numerical results show that the GBLinks model can reach a near-optimal solution through comparing with exhaustive search scheme in small-region ultra-dense D2D mmWave communication networks and outperforms GreedyNoSched and the SCA-based method. In addition, the GBLinks model has better generalization ability in terms of varying densities and coverage regions of ultra-dense D2D mmWave communication networks.
\end{itemize}


The rest of this paper is organized as follows. In Section \uppercase\expandafter{\romannumeral2}, we propose the spatial sharing D2D mmWave communication network and formulated joint beam selection and link activation problem as a binary integer programming non-convex optimization problem. In Section \uppercase\expandafter{\romannumeral3}, we solve it using the DC method and Lagrangian dual theory. In Section \uppercase\expandafter{\romannumeral4}, we propose a GNN-based model, i.e., GBLinks, to learn the joint beam selection and link activation policy. In Section \uppercase\expandafter{\romannumeral5}, we present the numerical results of the proposed method. Finally, we will conclude this paper in Section \uppercase\expandafter{\romannumeral6}. The main notations used throughout the
paper are summarized in Table \uppercase\expandafter{\romannumeral1}.

\begin{table}
  \renewcommand{\tablename}
  \caption{\centering{ TABLE \uppercase\expandafter{\romannumeral1}} \protect \\ \qquad \qquad \qquad \qquad \qquad  \textsc{List Of High Frequency Notations}}
  \\ \\
  \centering
\begin{tabular}{|c||c|c||c|} \hline
 \textbf{Notation} & \textbf{Description} & \textbf{Notation} & \textbf{Description} \\ \hline
 $\mathcal{N}$ &  The set of communication pairs & $N$ & Number of transmitter-receiver pairs \\ \hline
 $\phi_{m,r}$ & \makecell[c]{Receive analog beam indicator at \\ receiver $m$} & $\varphi_{n,l}$ & \makecell[c]{Transmit analog beam at transmitter $n$} \\ \hline
 $N_{\mathrm{t}}$ & \makecell[c]{The number of transmit antennas of \\ transmitter} & $N_{\mathrm{r}}$ & The number of transmit antennas of receiver \\ \hline
 $\mathcal{N}_{\mathrm{t}}$ & \makecell[c]{The index set of codewords of codebook $\mathcal{C}_{\mathrm{t}}$} & $\mathcal{N}_{\mathrm{r}}$ & \makecell[c]{The index set of codewords of codebook $\mathcal{C}_{\mathrm{r}}$} \\ \hline
 $\mathbf{\Psi}$ & \makecell[c]{The matrix of transmitting analog beam \\ indicators} & $\mathbf{\Phi}$ & \makecell[c]{The matrix of receiving analog beam \\ indicators} \\ \hline
 $\mathbf{U}_{\mathrm{t}},\mathbf{V}_{\mathrm{r}}$ & \makecell[c]{Codebook matrices for the transmitter \\and receiver, respectively} & $\mathbf{H}_{m,n}$ & \makecell[c]{Channel coefficient between the $m$-th \\ receiver and the $n$-th transmitter}
  \\ \hline
 $[\mathbf{A}]_{i,:}$ & The $i$-th row of matrix $\mathbf{A}$ & $[\mathbf{A}]_{i,j}$ & \makecell[c]{The element of the $i$-th row and\\ the $j$-th column of matrix $\mathbf{A}$} \\ \hline
 $|\cdot|$ & \makecell[c]{The absolute value of a complex \\ scalar or the cardinality of a set} & $\mathbb{C}$ & The set of complex numbers \\ \hline
 $\left(\cdot\right)^H$ & Hermite transpose & $\mathbb{R}_+$ & The set of positive real numbers \\ \hline
\end{tabular}
\end{table}


\section*{\sc \uppercase\expandafter{\romannumeral2}. System Model and Problem Formulation}

\subsection*{A. System Model}
\begin{figure}[ht]
\renewcommand{\captionfont}{\footnotesize}
\renewcommand*\captionlabeldelim{.}
	\centering
	\captionstyle{flushleft}
	\onelinecaptionstrue
	\includegraphics[width=0.8\columnwidth,keepaspectratio]{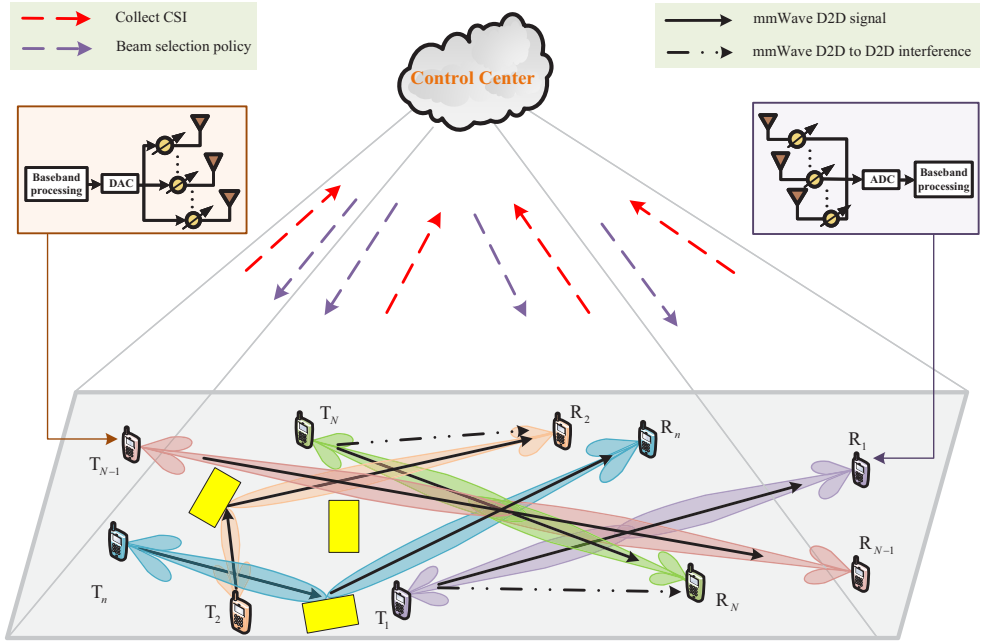}
	\caption{Illustration of spatial sharing mmWave communication.}
	\label{SystemModel}
\end{figure}

Consider an ultra-dense D2D mmWave communication network, as illustrated in Fig.~\ref{SystemModel}, in which there are $N$ distinct multi-antennas transmitter-receiver pairs to establish dense communication links via directional transmission on the same time-frequency resource. Let $\mathcal{N}=\left\{1,\cdots,N\right\}$ be the set of $N$ distinct multi-antennas transmitter-receiver pairs.
Each transmitter is equipped with a single RF chain connecting with $N_{\mathrm{t}}$ transmit antennas via $N_{\mathrm{t}}$ phase shifters~\footnote{It is worth noting that the proposed GBLinks model can be applied in the communication systems with an arbitrary number of RF chains by viewing each RF chain equipped in a transmitter or receiver as a virtual transmitter or receiver. Meanwhile, an indicator for each RF chain of a transmitter or receiver is used to determine whether a RF chain is selected or not. If all indicators for the RF chains of a transmitter or receiver are zeros, i.e., no RF chains are selected, the transmitter or receiver is not activated. Otherwise, the transmitter or receiver is activated.}. Similarly, each receiver is also equipped with a single RF chain connecting with $N_{\mathrm{r}}$ transmit antennas via $N_{\mathrm{r}}$ phase shifters. The $m$-th communication pair is consist of the $m$-th transmitter and the $m$-th receiver, $m\in\mathcal{N}$. Let $\mathbf{H}_{m,n}\in\mathbb{C}^{N_{\mathrm{r}}\times N_{\mathrm{t}}}$ denotes the channel coefficient between the $m$-th receiver and the $n$-th transmitter. For mmWave communication, generally speaking, there are only limited scatterers between communication pairs~\cite{JSTSPHeath2016}. Therefore, in this paper, channel matrix $\mathbf{H}_{m,n}$ is modeled as a narrowband clustered Saleh-Valenzuela model. Each scatterer is further assumed to contribute a single propagation path to the channel between the transmitter and receiver. Thus, channel matrix $\mathbf{H}_{m,n}$ is given by~\cite{el2014spatially}
\begin{equation}\label{GNNSPSH01}
\mathbf{H}_{m,n}=\sqrt{\rho_{m,n}N_{\mathrm{t}}N_{\mathrm{r}}}\sum\limits_{p=1}^{N_{\mathrm{p}}}
\alpha_{p,m,n}\mathbf{h}_{\mathrm{r}}\left(\tau_{p,m,n}\right)\mathbf{h}_{\mathrm{t}}^{H}\left(\psi_{p,m,n}\right),
\end{equation}
where $N_{\mathrm{p}}$ denotes the number of paths between communication pairs. $\tau_{p,m,n}\in\left[0,2\pi\right)$ and $\psi_{p,m,n}\in\left[0,2\pi\right)$ denote the azimuth angles of arrival and departure (AoA/AoD) of the $p$-th path between the $m$-th receiver and the $n$-th transmitter, respectively. $\rho_{m,n}$ and $\alpha_{p,m,n}$ denote respectively the average path-loss and the complex gain of the $p$-th path between the $m$-th receiver and the $n$-th transmitter. Assume that a uniform linear array (ULA) with half wavelength antenna spacing is adopted at the communication pairs. In particular, for an $N_{\mathrm{t}}$-element ULA, the array response vector is given by~\eqref{GNNSPSH02}. Similarly, $\mathbf{h}_{\mathrm{r}}\left(\tau_{p,m,n}\right)$  can be calculated.
\begin{equation}\label{GNNSPSH02}
\mathbf{h}_{\mathrm{t}}\left(\psi_{p,m,n}\right)=\sqrt{\frac{1}{N_{\mathrm{t}}}}
\left[1,e^{j\pi\sin\left(\psi_{p,m,n}\right)},\atop\cdots,
e^{j\left(N_{\mathrm{t}}-1\right)\pi\sin\left(\psi_{p,m,n}\right)}\right]^{T},
\end{equation}

Due to the existing large pathloss of mmWave communication, in general, analog beams adopted by the communication pairs need to be determined before formally transmitting data. One of the beam training methods is to train the analog beams based on a predefined codebook to obtain the optimum beam pairs that maximize the desired receiving signal energy without considering the interference between each other~\cite{JSACWang2009}. Similarly, in this paper, we assume that the analog beams used at the communication pairs come from a predesigned codebook. For ease of notation, let $\mathbf{U}_{\mathrm{t}}$ and $\mathbf{V}_{\mathrm{r}}$ be the codebook matrix used at the transmitter and receiver, respectively. The numbers of columns in codebook matrix $\mathbf{U}_{\mathrm{t}}$ and $\mathbf{V}_{\mathrm{r}}$ are $N_{\mathrm{t}}$ and $N_{\mathrm{r}}$, respectively. Furthermore, the column vector in $\mathbf{U}_{\mathrm{t}}$ and $\mathbf{U}_{\mathrm{r}}$ is the unit-norm vector. The baseband signal $y_{m}$ received at the $m$-th receiver is expressed as
\begin{equation}\label{Cachenable03}
y_{m}=\mathbf{v}_{m,r}^{H}\sum\limits_{n\in\mathcal{A}}\sqrt{p_{n}}\mathbf{H}_{m,n}\mathbf{u}_{n,t}x_{n}+\upsilon_{m},
\end{equation}
where $\mathbf{u}_{n,t}$ and $\mathbf{v}_{m,r}$ denote the $n$-th column of $\mathbf{U}_{\mathrm{t}}$ and the $m$-th column of $\mathbf{V}_{\mathrm{r}}$, respectively. $x_{n}$ is the transmitted signal at the $n$-th transmitter and $\upsilon_{m}\sim\mathcal{CN}\left(0,\sigma_{m}^{2}\right)$ is the additive white Gaussian noise. $p_{n}$ is the transmitting power of the $n$-th transmitter and $\mathcal{A}\subseteq\mathcal{N}$ is the set of the activated communication pairs.

\subsection*{B. Problem Formulation}
To maximize the total throughput of ultra-dense D2D mmWave communication networks, the interference between communication pairs needs be carefully controlled by properly selecting the analog transmitting and receiving beams and activating some communication links. In this subsection, we formulate the joint beam selection and link activation as maximizing the total throughput of ultra-dense D2D mmWave communication networks. To effectively characterize the beam selection, we define two indicators $\phi_{m,r}$ and $\varphi_{n,l}$ that denote the receiving and transmitting analog beam indicies used at the $m$-th receiver and the $n$-th transmitter, respectively. In particular, if the $n$-th transmitter adopts the $l$-th codeword, i.e., the $l$-th column of $\mathbf{U}_{\mathrm{t}}$, as the transmitting analog beam, then $\varphi_{n,l}=1$, otherwise $\varphi_{n,l}=0$, $n\in\mathcal{N}$, $l\in\mathcal{N}_{\mathrm{t}}=\{1,2,...,N_{\mathrm{t}}\}$. Similarly, if the $m$-th receiver uses the $r$-th codeword, i.e., the $r$-th column of $\mathbf{V}_{\mathrm{r}}$ as the receiving analog beam, then $\phi_{m,r}=1$, otherwise $\phi_{m,r}=0$, $m\in\mathcal{N}$, $r\in\mathcal{N}_{\mathrm{r}}=\{1,2,...,N_{\mathrm{r}}\}$. If $\varphi_{m,l}=0$, $\forall l\in\mathcal{N}_{\mathrm{t}}$ and $\phi_{m,r}=0$, $\forall r\in\mathcal{N}_{\mathrm{r}}$, then the $m$-th communication pair is inactivated, i.e., $m\notin\mathcal{A}$, otherwise, $m\in\mathcal{A}$. Further, for the considered ultra-dense D2D mmWave communication network, the indicators $\varphi_{n,t}$ and $\phi_{m,r}$ subject to the following constraints
\begin{subequations}\label{Cachenable04}
\begin{align}
&\varphi_{n,t}\in\left\{0,1\right\}, \forall n\in\mathcal{N}, t\in\mathcal{N}_{\mathrm{t}},\label{Cachenable04a}\\
&\phi_{m,r}\in\left\{0,1\right\}, \forall m\in\mathcal{N}, r\in\mathcal{N}_{\mathrm{r}},\label{Cachenable04b}\\
&\sum\limits_{t\in\mathcal{N}_{\mathrm{t}}}\varphi_{n,t}\leqslant 1, \forall n\in\mathcal{N},\label{Cachenable04c} \\
&\sum\limits_{r\in\mathcal{N}_{\mathrm{r}}}\phi_{m,r}\leqslant 1, \forall m\in\mathcal{N},\label{Cachenable04d}\\
&\sum\limits_{t\in\mathcal{N}_{\mathrm{t}}}\varphi_{m,t}=\sum\limits_{r\in\mathcal{N}_{\mathrm{r}}}\phi_{m,r}, \forall m\in\mathcal{N}.\label{Cachenable04e}
\end{align}
\end{subequations}
Constraints~\eqref{Cachenable04a} and~\eqref{Cachenable04b} make $\varphi_{n,t}$ and $\phi_{m,r}$ be binary variables. Constraints~\eqref{Cachenable04c} and~\eqref{Cachenable04d} assure that the transmitter and receiver only select a single beam for each communication link. Constraint~\eqref{Cachenable04e} assures that the receiving and transmitting beams are simultaneously activated for a communication pair. Thus, without introducing confusion, the achievable rate $R_{m,r,t}$ of the $m$-th communication pair with the $r$-th receiving beam at the $m$-th receiver and the $t$-th transmitting beam at the $m$-th transmitter is defined as
\begin{equation}\label{Cachenable05}
R_{m,r,t}=\log_{2}\left(1+\frac{\phi_{m,r}\varphi_{m,t}p_{m}\varrho\left(m,r,m,t\right)}
{\sum\limits_{n\in\mathcal{N}\setminus\left\{m\right\}}\sum\limits_{l\in\mathcal{N}_{\mathrm{t}}}\phi_{m,r}\varphi_{n,l}p_{n}\varrho\left(m,r,n,l\right)+\sigma_{m}^{2}}\right),
\end{equation}
where $\varrho\left(m,r,n,l\right)=\left|\mathbf{v}_{m,r}^{H}\mathbf{H}_{m,n}\mathbf{u}_{n,l}\right|^{2}$, $\sigma_m^2$ denotes the noise variance of the $m$-th communication pair. It is worth noting that the achievable rate $R_{m,r,t}$ is defined under the constraints~\eqref{Cachenable04a}-\eqref{Cachenable04e}, which guarantees each transmitter or receiver only uses a single beam or do not use beam from the system model. For ease of presentation, let $\bm{\Psi}$ and $\bm{\Phi}$ be the indicator matrices for the transmitting and receiving analog beam, respectively, where $\left[\bm{\Psi}\right]_{n,l}=\varphi_{n,l}$ and $\left[\bm{\Phi}\right]_{m,r}=\phi_{m,r}$. The corresponding optimization problem is formulated as
\begin{equation}\label{Cachenable06}
\max\limits_{\bm{\Psi},\bm{\Phi}}\sum_{m\in\mathcal{N}}\sum\limits_{r\in\mathcal{N}_{\mathrm{r}}}\sum\limits_{t\in\mathcal{N}_{\mathrm{t}}}w_mR_{m,r,t},
~\mathrm{s.t.}~\textrm{(4a)}-\textrm{(4e)}.
\end{equation}
where $w_m$ is the weight coefficient of the $m$-th communication pair. The goal of the problem~\eqref{Cachenable06} is to activate as more communication pairs as possible aiming at maximizing the throughput of ultra-dense D2D mmWave communication networks. As we known that the user rate function is non-convex, therefore, problem~\eqref{Cachenable06} is a binary integer programming non-convex optimization problem, which is very difficult to solve.

As far as we know, our work is the first to perform joint beam selection and link activation for ultra-dense D2D mmWave communication networks with multiple communication pairs. Generally speaking, there are two simple scheme to solve problem~\eqref{Cachenable06}, which are illustrated as follows

\begin{itemize}
  \item \textbf{Exhaustive Search Scheme}: There is no doubt that the optimal solution can be obtained using exhaustive search in the beam space of communication pairs. However, the computational complexity of exhaustive search scheme is $\mathcal{O}(\sum_{k=1}^{N}C_N^k(N_\mathrm{r} N_\mathrm{t})^k)$ where $C_N^k=\frac{N!}{k!(N-k)!}$. The computational overhead of exhaustive search scheme will rise sharply as the numbers of beams and communication pairs increase.
  \item \textbf{GreedyNoSched}: There is a simple method without considering the interference between communication pairs. Specifically, each communication pair chooses the beam pair that maximizes the desired receiving signal energy, we call it GreedyNoSched scheme. The computational complexity of GreedyNoSched is $\mathcal{O}(NN_\mathrm{r}N_\mathrm{t})$. Although the complexity of this scheme is low, it may not be ideal in performance as illustrated in Section V.
\end{itemize}

Due to the drawback of exhaustive search scheme and GreedyNosched, an effective scheme of joint beam selection and link activation is needed to improve the performance in terms of the weighted sum rate with a proper computational complexity. Note that the difficulties of solving problem~\eqref{Cachenable05} are the binary optimization variable and non-convex objective function. To obtain a tractable form of problem~\eqref{Cachenable05}, constraints~\eqref{Cachenable04a} and~\eqref{Cachenable04b} are equivalently reformulated as follows~\cite{he2017joint}
\begin{subequations}\label{Cachenable07}
\begin{align}
&0\leq\varphi_{n,t}\leq1, \forall n\in\mathcal{N}, t\in\mathcal{N}_{\mathrm{t}},\label{Cachenable07a}\\
&0\leq\phi_{m,r}\leq1, \forall m\in\mathcal{N}, r\in\mathcal{N}_{\mathrm{r}},\label{Cachenable07b}\\
&\varphi_{n,t} - \varphi^2_{n,t}\leq0, \forall n\in\mathcal{N}, t\in\mathcal{N}_{\mathrm{t}},\label{Cachenable07c}\\
&\phi_{m,r} - \phi^2_{m,r}\leq0, \forall m\in\mathcal{N}, r\in\mathcal{N}_{\mathrm{r}}.\label{Cachenable07d}
\end{align}
\end{subequations}
In this way, variables $\phi_{m,r}$ and $\varphi_{n,t}$ are continuous values between 0 and 1 while inequalities~\eqref{Cachenable07c} and~\eqref{Cachenable07d} assure that the values are zero or one. Thus, problem~\eqref{Cachenable05} can be rewritten as
\begin{equation}\label{Cachenable08}
\max\limits_{\bm{\Psi},\bm{\Phi}}\sum_{m\in\mathcal{N}}\sum\limits_{r\in\mathcal{N}_{\mathrm{r}}}\sum\limits_{t\in\mathcal{N}_{\mathrm{t}}}w_mR_{m,r,t}, \\
~\mathrm{s.t.}~\textrm{(4c)}-\textrm{(4e)}, \textrm{(7a)}-\textrm{(7d)}.
\end{equation}
In general, problem~\eqref{Cachenable08} is a non-convex and NP-hard problem. Its optimal solution is difficult to obtain, even the local optimal solution cannot be obtained directly. Although optimization problem~\eqref{Cachenable08} can be solved by the SCA method~\cite{SCA2015}, which is proposed in the Appendix, it requires a lot of computational overhead.

The proposed SCA-based method has several defects, which are not limited to the SCA-based method. Although the SCA-based method can get a local optimal solution, its computational complexity increases sharply with the increase of the number of communication pairs and antennas. Furthermore, the SCA-based method not only needs to solve problem~\eqref{Cachenable08} for each single instance, but also may need to adjust the parameters of the SCA-based method for each specific instance.  Motivated by these observations, we would like to resort an effective and efficient manner to solve problem~\eqref{Cachenable08}. Specifically, inspired by the successful application of GNN in wireless networks~\cite{shen2019graph}, we propose to utilize GNN to learn the mapping between the channel coefficient matrix $\mathbf{H}$, codebook matrices $\mathbf{U}_{\mathrm{t}}$ and $\mathbf{V}_{\mathrm{r}}$, and the corresponding~(near)-optimal beam selection policies $\mathbf{\Psi}$ and $\mathbf{\Phi}$. In this way, the beam selection policies can be efficiently obtained by the learned mapping without solving the optimization problem for each new instance. In the sequel, we elaborate the design of GNN-based model, i.e., GBLinks, and its training scheme to address problem~\eqref{Cachenable08}.

\section*{\sc \uppercase\expandafter{\romannumeral3}. Lagrangian Dual Learning Framework}

In this section, we focus on proposing a Lagrangian Dual Learning Framework (LDLF) to train the GBLinks model designed and discussed in detail in the following section. Due to there are a large mount of constraints in problem~\eqref{Cachenable08}, it is firstly transformed into an unconstrained optimization problem. Specifically, the Lagrange multiplier method is utilized to deal with problem~\eqref{Cachenable08}. Let $\bm{\nu},\bm{\xi},\bm{\rho}\in \mathbb{R}_{+}^{N}$, and $\bm{\lambda}\in \mathbb{R}_{+}^{N\times N_\mathrm{t}}$, $\bm{\mu}\in \mathbb{R}_{+}^{N\times N_\mathrm{r}}$, be the nonnegative dual multipliers associated with constraints~\eqref{Cachenable04c}-\eqref{Cachenable04e} and~\eqref{Cachenable07c}-\eqref{Cachenable07d}, respectively. In order to capture how much the constraints are violated and to guarantee the update value of Lagrangian multipliers is always positive, we use the violation degree for~\eqref{Cachenable04c}-\eqref{Cachenable04e}, and relax~\eqref{Cachenable07c} and~\eqref{Cachenable07d} based on the satisfiability degree. Then, the partial Lagrangian relaxation function of problem~\eqref{Cachenable08} is formulated as
\begin{equation}\label{Cachenable09}
\begin{aligned}
&\mathcal{L}\left(\bm{\Phi},\bm{\Psi},\bm{\lambda},\bm{\mu},\bm{\nu},\bm{\xi},\bm{\rho}\right)=-\sum_{m\in\mathcal{N}}\sum\limits_{r\in\mathcal{N}_{\mathrm{r}}}\sum\limits_{t\in\mathcal{N}_{\mathrm{t}}}w_mR_{m,r,t}+\sum\limits_{m\in\mathcal{N}}\sum\limits_{t\in\mathcal{N}_\mathrm{t}}\bm{\lambda}_{m,t}\sigma_c\left(\varphi_{m,t},\varphi_{m,t}^2\right) \\
&~~~~~+\sum\limits_{m\in\mathcal{N}}\sum\limits_{r\in\mathcal{N}_\mathrm{r}}\bm{\mu}_{m,r}\sigma_c\left(\phi_{m,r},\phi_{m,r}^2\right)+\sum\limits_{m\in\mathcal{N}}\bm{\nu}_{m}\chi_c^{\leq}\left(\sum\limits_{t\in\mathcal{N}_\mathrm{t}}\varphi_{m,t}\right) +\sum\limits_{m\in\mathcal{N}}\bm{\xi}_{m}\chi_c^{\leq}\left(\sum\limits_{r\in\mathcal{N}_\mathrm{r}}\phi_{m,r}\right) \\
&~~~~~+\sum\limits_{m\in\mathcal{N}}\bm{\rho}_m\chi_c^{=}\left(\sum\limits_{t\in\mathcal{N}_\mathrm{t}}\varphi_{m,t},\sum\limits_{r\in\mathcal{N}_\mathrm{r}}\phi_{m,r}\right),
\end{aligned}
\end{equation}
where $\chi_c^{\leq}(x_1,...,x_n)=\mathrm{max}\left(0,\sigma_c(x_1,...,x_n)\right)$ is the violation degree for inequality constraints like $c(x_1,...,x_n)\equiv\sigma_c(x_1,...,x_n)\leq0$, $\chi_c^{=}(x_1,...,x_n)=|\sigma_c(x_1,...,x_n)|$ is the violation degree of equality constraints like $c(x_1,...,x_n)\equiv\sigma_c(x_1,...,x_n)=0$, and $\sigma_c(x_1,...,x_n)$ is a function $\mathbb{R}^n\rightarrow \mathbb{R}$. Further, the Lagrangian dual optimization problem is formulated as
\begin{equation}\label{Cachenable10}
\max\limits_{\bm{\lambda},\bm{\mu},\bm{\nu},\bm{\xi},\bm{\rho}}~\min\limits_{\bm{\Phi},\bm{\Psi}}\mathcal{L}\left(\bm{\Phi},\bm{\Psi},\bm{\lambda},\bm{\mu},\bm{\nu},\bm{\xi},\bm{\rho}\right).
\end{equation}

In general, alternative optimization is a preferable selection for solving the max-min optimization problem, as shown in Fig.~2. In particular, we minimize Lagrangian relaxation function~\eqref{Cachenable09} over primal variables $\bm{\Phi}$ and $\bm{\Psi}$ with fixed dual variables. Then, we maximize Lagrangian relaxation function~\eqref{Cachenable09} over dual variables $\bm{\lambda},\bm{\mu},\bm{\nu},\bm{\xi}$ and $\bm{\rho}$ with fixed primal variables.
For the outer optimization problem, taking the $\ell$-th iteration as example with given primal variables $\bm{\Phi}^{(\ell-1)}$ and $\bm{\Psi}^{(\ell-1)}$, the Lagrangian multipliers are updated using the subgradient method, i.e.,
\begin{subequations}\label{Cachenable11}
\begin{align}
&\bm{\lambda}_{m,t}^{(\ell)}=\bm{\lambda}_{m,t}^{(\ell-1)}+\varepsilon_{\lambda}\sigma_{c}\left(\varphi_{m,t}^{(\ell-1)},\left(\varphi_{m,t}^{(\ell-1)}\right)^2\right),\forall m\in\mathcal{N},t\in\mathcal{N}_{\mathrm{t}},\label{Cachenable11a}\\
&\bm{\mu}_{m,r}^{(\ell)}=\bm{\mu}_{m,r}^{(\ell-1)}+\varepsilon_{\mu}\sigma_{c}\left(\phi_{m,r}^{(\ell-1)},\left(\phi_{m,r}^{(\ell-1)}\right)^2\right),\forall m\in\mathcal{N}, r\in\mathcal{N}_{\mathrm{r}},\label{Cachenable11b}\\
&\bm{\nu}_m^{(\ell)}=\bm{\nu}_m^{(\ell-1)}+\varepsilon_{\nu}\chi_{c}^{\leq}\left(\sum\limits_{t\in\mathcal{N}_{\mathrm{t}}}\varphi_{m,t}^{(\ell-1)}\right),\forall m\in\mathcal{N},\label{Cachenable11c}\\
&\bm{\xi}_m^{(\ell)}=\bm{\xi}_m^{(\ell-1)}+\varepsilon_{\xi}\chi_{c}^{\leq}\left(\sum\limits_{r\in\mathcal{N}_{\mathrm{r}}}\phi_{m,r}^{(\ell-1)}\right),\forall m\in\mathcal{N},\label{Cachenable11d}\\
&\bm{\rho}_m^{(\ell)}=\bm{\rho}_m^{(\ell-1)}+\varepsilon_{\rho}\chi_c^{=}\left(\sum\limits_{t\in\mathcal{N}_\mathrm{t}}\varphi_{m,t}^{(\ell-1)},\sum\limits_{r\in\mathcal{N}_\mathrm{r}}\phi_{m,r}^{(\ell-1)}\right),\forall m\in\mathcal{N},\label{Cachenable11e}
\end{align}
\end{subequations}
where $\varepsilon_{\lambda},\varepsilon_{\mu},\varepsilon_{\nu},\varepsilon_{\xi}$, and $\varepsilon_{\rho}>0$ denote the update step-sizes of Lagrangian multipliers $\bm{\lambda},\bm{\mu},\bm{\nu},\bm{\xi}$, and $\bm{\rho}$, respectively. Instead of using directly the traditional optimization method to solve the inner optimization problem, in what follows, we focus on designing the GBLinks model, to optimize primal parameters $\bm{\Phi}^{(\ell)}$ and $\bm{\Psi}^{(\ell)}$ with given Lagrangian multipliers $\bm{\lambda}^{(\ell-1)},\bm{\mu}^{(\ell-1)},\bm{\nu}^{(\ell-1)},\bm{\xi}^{(\ell-1)}$, and $\bm{\rho}^{(\ell-1)}$ in the $\ell$-th iteration. In other words, the GBLinks model is trained to minimize Lagrangian relaxation function~\eqref{Cachenable09} with the fixed Lagrangian multipliers, i.e.,
\begin{equation}\label{Cachenable12}
\begin{aligned}
\mathcal{L}\left(\bm{\Phi},\bm{\Psi},\bm{\lambda}^{(\ell-1)},\bm{\mu}^{(\ell-1)},\bm{\nu}^{(\ell-1)},\bm{\xi}^{(\ell-1)},\bm{\rho}^{(\ell-1)}\right), \end{aligned}
\end{equation}
which is defined as the loss function of the GBLinks model in the $\ell$-th iteration. In the sequel, we focus on  describing the constructions of the GBLinks model.
\begin{figure}[t]
\renewcommand{\captionfont}{\footnotesize}
\renewcommand*\captionlabeldelim{.}
	\centering
	\captionstyle{flushleft}
	\onelinecaptionstrue
	\includegraphics[width=0.8\columnwidth,keepaspectratio]{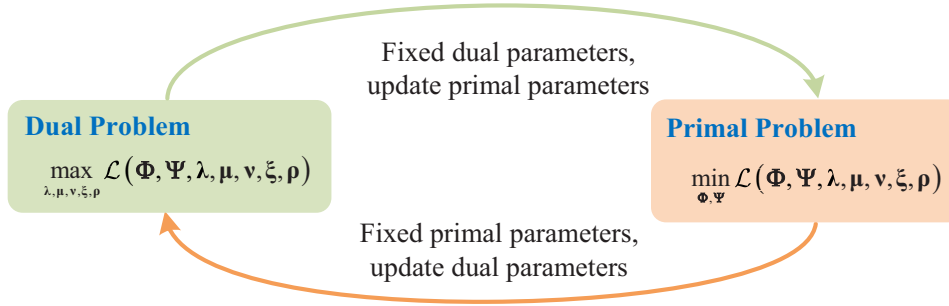}
	\caption{Illustration of Lagrangian dual learning framework.}
	\label{LDF1}
\end{figure}

\section*{\sc \uppercase\expandafter{\romannumeral4}. Design Of The GBLinks Model}
In this section, we focus on designing the GBLinks model to learn the policies of beam selection and link activation in an unsupervised manner, which is under the framework of GNNs. We first introduce the method of building wireless channel graph for ultra-dense D2D mmWave communication networks, and then propose a GBLinks model to learn the joint beam selection and link activation policies.

\subsection*{A. Wireless Channel Graph Construction}
\begin{figure}[t]
\renewcommand{\captionfont}{\footnotesize}
\renewcommand*\captionlabeldelim{.}
	\centering
	\captionstyle{flushleft}
	\onelinecaptionstrue
	\includegraphics[width=0.8\columnwidth,keepaspectratio]{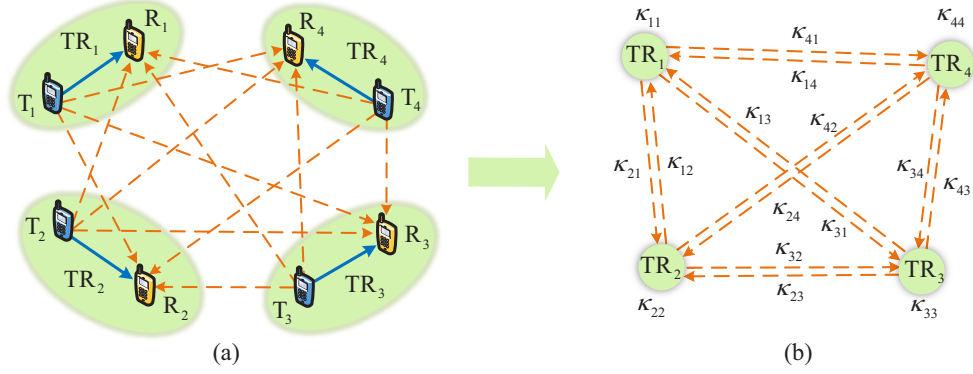}
	\caption{Four communication pairs and the corresponding wireless channel graph.}
	\label{graphbuild}
\end{figure}

In this subsection, we focus on building a wireless channel graph for the considered ultra-dense D2D mmWave communication network with $N$ distinct multi-antennas communication pairs. Note that a communication pair is activated or deactivated and may interfere with each other in the ultra-dense D2D mmWave communication network. The wireless channel graph is constructed as a directed
complete wireless channel graph, i.e., $\mathcal{G}\left(\mathcal{V,E}\right)$, where $\mathcal{V}$ is the vertex set consisting of all the communication pairs, and $\mathcal{E}$ is the edge set including the interference links between different communication pairs. Fig.~\ref{graphbuild}. gives out an example of a directed wireless channel graph for a four communication pairs D2D mmWave communication network. In Fig.~\ref{graphbuild}(a), $\mathrm{T}_i$ and $\mathrm{R}_i$ represent the $i$-th transmitter and the $i$-th receiver, respectively, and $\mathrm{TR}_i$ represents the $i$-th communication pair, $i\in\mathcal{N}$. The blue arrow denotes the direct link and the yellow arrow denotes the interference link. As shown in Fig.~\ref{graphbuild}(b), the green vertex represents each communication pair $\mathrm{TR}_i,i\in\mathcal{N}$. To better describe the wireless channel graph, we firstly define some effective information features for each vertex and each edge~\footnote{Generally speaking, the acquisition cost of physical CSI is large in multi-antenna wireless communication systems. Note that the proposed GBLinks model only depends on the equivalent CSI, which can be obtained by beam scanning without needing the specific training overhead for the acquisition of physical CSI. Therefore, the proposed GBLinks model is effective and efficient for the ultra-dense D2D mmWave communication networks. In the future work, to efficiently handle the ultra-dense D2D mmWave communication networks with massive communication pairs, we would like to investigate a distributed learning model for ultra-dense D2D mmWave communication networks.}. Specifically, the features of all vertices and edges are denoted as a tensor $\bm{\kappa}\in\mathbb{R}^{N\times{N}\times{d}}$. $\bm{\kappa}_{i,i}\in{\mathbb{R}^d}$ and $\bm{\kappa}_{i,j}\in{\mathbb{R}^d}$ represent the feature vectors of the $i$-th vertex and that of the directed edge from vertex $i$ to vertex $j$, $i\in\mathcal{N}, j\in\mathcal{N}\setminus\{i\}$, with $d=N_{\mathrm{t}}N_{\mathrm{r}}$ being the feature dimension of both vertices and edges. The feature vector of the $i$-th vertex is defined as  $\bm{\kappa}_{i,i}=g\left(f\left(\mathbf{V}_\mathrm{r}^{H}\mathbf{H}_{i,i}\mathbf{U}_\mathrm{t}\right)\right),i\in\mathcal{N}$, where $f(\cdot):~\mathbb{C}^{N_\mathrm{r}\times N_\mathrm{t}}\rightarrow\mathbb{C}^d$, represents the operation of flattening elements in a matrix row-by-row,  $g(\cdot):~\mathbb{C}^d\rightarrow\mathbb{R}^d$, denotes the element-wise operation of taking the modulus of complex. For example, suppose $\hat{\bm{\kappa}}_{i,i}=\mathbf{V}_{\mathrm{r}}^{H}\mathbf{H}_{i,i}\mathbf{U}_{\mathrm{t}}=[a_{1,1},...,a_{1,N_{\mathrm{t}}};...;a_{N_{\mathrm{r}},1},...,a_{N_{\mathrm{r}},N_{\mathrm{t}}}]\in\mathbb{C}^{N_{\mathrm{r}}\times N_{\mathrm{t}}}$, then $f(\hat{\bm{\kappa}}_{i,i})=[a_{1,1},...,a_{1,N_{\mathrm{t}}},...,a_{N_{\mathrm{r}},1},...,a_{N_{\mathrm{r}},N_{\mathrm{t}}}]^T\in\mathbb{C}^{d}$, and $\bm{\kappa}_{i,i}=g\left(f(\hat{\bm{\kappa}}_{i,i})\right)=[|a_{1,1}|,...,|a_{1,N_{\mathrm{t}}}|,...,|a_{N_{\mathrm{r}},1}|,...,|a_{N_{\mathrm{r}},N_{\mathrm{t}}}|]^T\in\mathbb{R}^{d}$. The edges between two vertices are directed, indicating the interference links of two vertices and the feature vectors of edges between vertex $i$ and vertex $j$ are defined as $\bm{\kappa}_{i,j}=f\left(\mathbf{V}_{\mathrm{r}}^{H}\mathbf{H}_{i,j}\mathbf{U}_{\mathrm{t}}\right) ,\bm{\kappa}_{j,i}=f\left(\mathbf{V}_{\mathrm{r}}^{H}\mathbf{H}_{j,i}\mathbf{U}_{\mathrm{t}}\right),i\in\mathcal{N},j\in\mathcal{N}\setminus\left\{i\right\}$.

\subsection*{B. Implementation of the GBLinks Model}
\begin{figure}[ht]
\renewcommand{\captionfont}{\footnotesize}
\renewcommand*\captionlabeldelim{.}
	\centering
	\captionstyle{flushleft}
	\onelinecaptionstrue
	\includegraphics[width=0.9\columnwidth,keepaspectratio]{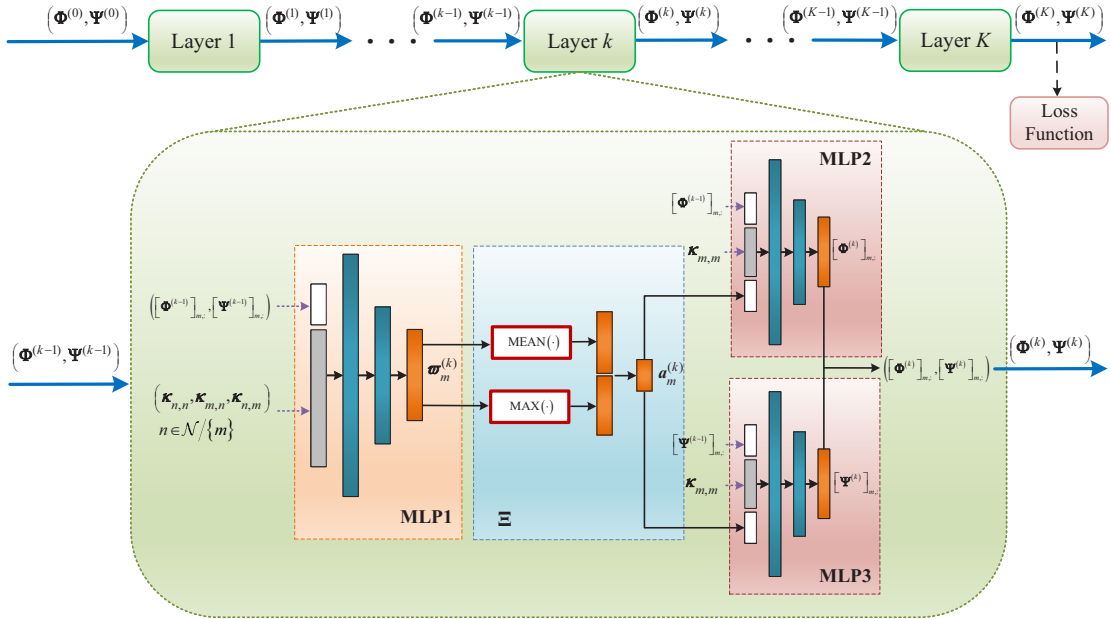}
	\caption{A illustration of GBLinks with $K$ layers.
}
	\label{GBLinks-v1}
\end{figure}

The GBLinks model is an end-to-end learning model and is consist of $K$ layers, as shown in Fig.~\ref{GBLinks-v1}. Each layer consists of a graph convolution module. Let $\bm{\Phi}^{(k)}\in\mathbb{R}^{N\times N_\mathrm{r}}$ and $\bm{\Psi}^{(k)}\in\mathbb{R}^{N\times N_\mathrm{t}}$ be the beam selection policies of receiver and transmitter obtained in the $k$-th layer, respectively. The input of the GBLinks model is $\left(\bm{\Phi}^{(0)},\bm{\Psi}^{(0)}\right)$, and the final beam selection policies are $\left(\bm{\Phi}^{(K)},\bm{\Psi}^{(K)}\right)$ outputted by the $K$-th layer. For the convenience of description, we only describe the update mode of the beam selection policies of the $m$-th vertex in the $k$-th layer. The input of the $k$-th layer is the output of the $(k-1)$-th layer, i.e., $\left(\bm{\Phi}^{(k-1)},\bm{\Psi}^{(k-1)}\right)$. The dashed boxes identified as $\mathrm{MLP1},\mathrm{\Xi},\mathrm{MLP2}$, and $\mathrm{MLP3}$ represent the main functional modules of graph convolution module. Specifically, $\mathrm{MLP1}$ is a MLP for aggregating the information of the beam selection and the features of neighbor vertices and edges. $\mathrm{\Xi}$ is a function to generate a vector. $\mathrm{MLP2}$ and $\mathrm{MLP3}$ are used to combine the aggregated information, and to update the beam selection policies of communication pairs. The white boxes and grey boxes correspond to dynamic inputs and the static input tensor $\bm{\kappa}$, respectively. The blue boxes denote the hidden layers of MLPs and the orange boxes denote the output of each functional module.

\begin{figure}[t]
\renewcommand{\captionfont}{\footnotesize}
\renewcommand*\captionlabeldelim{.}
	\centering
	\captionstyle{flushleft}
	\onelinecaptionstrue
	\includegraphics[width=0.8\columnwidth,keepaspectratio]{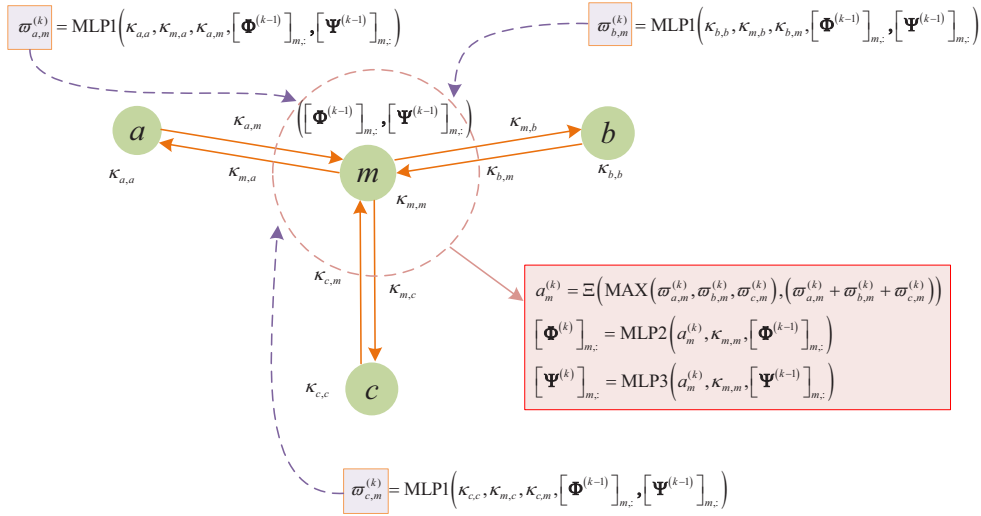}
	\caption{Illustration of implementating graph convolution module.}
	\label{gcn-m}
\end{figure}

The core of the GBLinks model is to design the graph convolution module, which is utilized to pass and update the information of vertices or edges of the directed complete wireless channel graph. The information passing and updating mechanisms are called AGGREGATE and COMBINE, respectively, which are the most important functions we should discuss in the follows. Generally speaking, the functions AGGREGATE and COMBINE are used to update a vertex's hidden state, i.e.,  the beam selection policies. The information aggregation and combination strategies AGGREGATE and COMBINE are designed using the spatial-based graph convolution network. At the $k$-th layer, the AGGREGATE mechanism is designed as
\begin{subequations}\label{Cachenable13}
\begin{align}
&\bm{\varpi}_{n,m}^{(k)}=\mathrm{MLP1}\left(\bm{\kappa}_{n,n},\bm{\kappa}_{m,n},\bm{\kappa}_{n,m},[\bm{\Phi}^{(k-1)}]_{m,:}, [\bm{\Psi}^{(k-1)}]_{m,:}\right),n\in\mathcal{N}/\{m\}, \label{Cachenable13a}\\
&\bm{a}_{m}^{(k)}=\Xi\left(\mathrm{MAX}\left(\bm{\varpi}_{n,m}^{(k)},n\in\mathcal{N}/\{m\}\right), \mathrm{MEAN}\left(\bm{\varpi}_{n,m}^{(k)},n\in\mathcal{N}/\{m\}\right)\right), \label{Cachenable13b}
\end{align}
\end{subequations}
where $\bm{\varpi}_{n,m}^{(k)}\in\mathbb{R}^{f}$ represents the information aggregated from vertex $n$ to vertex $m$ in the $k$-th layer, with $f$ being the output dimension of $\mathrm{MLP1}$. Function $\mathrm{MAX}(\cdot)$ is to take the largest value in a set in element-wise. Function $\mathrm{MEAN}(\cdot)$ is to take the mean value of a vector set in element-wise. The two symmetric functions $\mathrm{MAX}(\cdot)$ and $\mathrm{MEAN}(\cdot)$ are the key functions that guarantee the permutation invariance. $\bm{a}_{m}^{(k)}\in\mathbb{R}^{2f}$ is the information aggregated from all the neighbors of vertex $m$ in the $k$-th layer. In addition, $\mathrm{MLP1}$ uses ReLU~(Rectified Linear Unit) activation function, i.e., $\max\left(x,0\right)$, in the hidden and the final layer. While the COMBINE mechanism, namely, the update of beam selection policies $\left([\bm{\Phi}^{(k-1)}]_{m,:},[\bm{\Psi}^{(k-1)}]_{m,:}\right)$ of vertex $m$ is defined as
\begin{subequations}\label{Cachenable14}
\begin{align}
&[\bm{\Phi}^{(k)}]_{m,:}=\mathrm{MLP2}\left(\bm{a}_{m}^{(k)},\bm{\kappa}_{m,m},[\bm{\Phi}^{(k-1)}]_{m,:}\right),\label{Cachenable14a}\\
&[\bm{\Psi}^{(k)}]_{m,:}=\mathrm{MLP3}\left(\bm{a}_{m}^{(k)},\bm{\kappa}_{m,m},[\bm{\Psi}^{(k-1)}]_{m,:}\right),\label{Cachenable14b}
\end{align}
\end{subequations}
where $\mathrm{MLP2}$ and $\mathrm{MLP3}$ are designed as two different MLPs. In the final layer of $\mathrm{MLP2}$ and $\mathrm{MLP3}$, we use element-wise projection activation function $\Omega(\cdot)$ to project $\bm{\Phi}$ and $\bm{\Psi}$ onto the feasible region $\mathcal{O}\triangleq\{\bm{\Phi},\bm{\Psi}: 0\leq[\bm{\Phi}]_{m,r},[\bm{\Psi}]_{m,t}\leq1,\forall m\in\mathcal{N},r\in\mathcal{N}_\mathrm{r},t\in\mathcal{N}_\mathrm{t}\}$. In particular, the projection activation function $\Omega(\cdot)$ is defined as $\Omega(u)=\mathrm{max}\{0, \mathrm{min}\{u,1\}\}$, where $u$ is the variable should be activated. In this way, the final output of $\bm{\Phi}$ and $\bm{\Psi}$ will be projected to $[0,1]$. For ease of description, we give an illustration of implementing such graph convolution module for vertex $m$, which is shown in Fig.~\ref{gcn-m}. Vertex $m$ and its neighbor vertices, i.e., $a,b$ and $c$, interfere with each other. Before aggregating the feature information from vertex $m$'s neighbor vertices, each neighbor vertex will firstly do a nonlinear transform for vertex feature, edge feature and the beam selection polices outputted by the previous iteration. Then, vertex $m$ aggregates the transformed neighbor vertices' information and updates its beam selection policies. In the next subsection, we will focus on proposing a learning strategy for GBLinks model.

\begin{remark}
The proposed GBLinks model is an unsupervised learning model without requiring the ground truth, which just needs the feature $\bm{\kappa}$ as input. The loss function associated with the GBLinks model is Lagrangian relaxation function~\eqref{Cachenable12} with fixed Lagrangian multipliers. It is worth noting that equations~\eqref{Cachenable13} and~\eqref{Cachenable14} are vertex level operations. Therefore, the input of the GBLinks model is only limited to the dimension of the features, i.e., the number of antennas, not to the number of communication pairs. Further, the input sequence of communication pairs does not need to be fixed because two symmetric functions $\mathrm{MAX}\left(\cdot\right)$ and $\mathrm{MEAN}\left(\cdot\right)$ are used~\cite{unfolding2021}. In addition, the output of the GBLinks model can be adjusted adaptively according to the number of communication pairs. As long as the antenna configuration of the ultra-dense D2D mmWave communication networks remains unchange, the application of the GBLinks model will not be affected. Therefore, the GBLinks model adapts to a network with varying number of communication pairs with the same configuration of antennas.
\end{remark}

\begin{algorithm}[!ht]
\caption{LDLF for Constrained Problems}
\label{alg:algorithm1}
\begin{algorithmic}[1]
\renewcommand{\algorithmicrequire}{\textbf{Input:}}
\REQUIRE
$\mathcal{D}=\{\bm{\kappa}^{(l)}\}_{l=1}^n$: Training data \\
~~~~$\zeta,\varepsilon_{\lambda}, \varepsilon_{\mu}, \varepsilon_{\nu}, \varepsilon_{\xi}, \varepsilon_{\rho}$:~Optimizer and Lagrangian step sizes, respectively.
\STATE $\bm{\lambda}^{(0)},\bm{\mu}^{(0)},\bm{\nu}^{(0)},\bm{\xi}^{(0)},\bm{\rho}^{(0)} \gets 0$ \\
\FOR{\textit{epoch} $e\gets0,1,...$}
\STATE $\nabla_{\bm{\lambda}}^{(dual)},\nabla_{\bm{\mu}}^{(dual)},\nabla_{\bm{\nu}}^{(dual)},\nabla_{\bm{\xi}}^{(dual)},\nabla_{\bm{\rho}}^{(dual)}\gets0$
\STATE \textbf{for each} $\mathcal{B}_b \gets minibatch\left(\mathcal{D}\right)$ \textit{of size} $b$ \textbf{do}
\STATE~~~Obtain beam selection policies of a batch:  $\{\bm{\Psi}_{(i)}\}_{i=1}^{b},\{\bm{\Phi}_{(i)}\}_{i=1}^{b}\gets \text{GBLinks}\gets\mathcal{B}_b$ \\
\STATE~~~Batch-wise Lagrangian relaxation loss: \\ ~~~~~~~~~~~~~~$\mathcal{L}_b\gets\sum_{i=1}^{b}\mathcal{L}\left(\bm{\Psi}_{(i)},\bm{\Phi}_{(i)},\bm{\lambda}^{(e)},\bm{\mu}^{(e)},\bm{\nu}^{(e)},\bm{\xi}^{(e)},\bm{\rho}^{(e)}\right)$ \\
\STATE~~~Update the parameters of GBLinks: $\bm{w}\gets\bm{w}-\zeta\nabla_{\bm{w}}\mathcal{L}_b$ \\
\STATE~~~ $\nabla_{\bm{\lambda}_{m,t}}^{(dual)}\gets\nabla_{\bm{\lambda}_{m,t}}^{(dual)}+\sum_{i=1}^{b}\sigma_c\left([\bm{\Psi}_{(i)}]_{m,t},[\bm{\Psi}_{(i)}]_{m,t}^2\right),\forall m\in\mathcal{N},t\in\mathcal{N}_{\mathrm{t}}$
\STATE~~~ $\nabla_{\bm{\mu}_{m,r}}^{(dual)}\gets\nabla_{\bm{\mu}_{m,r}}^{(dual)}+\sum_{i=1}^{b}\sigma_c\left([\bm{\Phi}_{(i)}]_{m,r},[\bm{\Phi}_{(i)}]_{m,r}^2\right),\forall m\in\mathcal{N},t\in\mathcal{N}_{\mathrm{r}}$
\STATE~~~ $\nabla_{\bm{\nu}_m}^{(dual)}\gets\nabla_{\bm{\nu}_m}^{(dual)}+\sum_{i=1}^{b}\chi_c^{\leq}\left(\sum_{t\in\mathcal{N}_{\mathrm{t}}}[\bm{\Psi}_{(i)}]_{m,t}\right),\forall m\in\mathcal{N}$
\STATE~~~ $\nabla_{\bm{\xi}_m}^{(dual)}\gets\nabla_{\bm{\xi}_m}^{(dual)}+\sum_{i=1}^{b}\chi_c^{\leq}\left(\sum_{r\in\mathcal{N}_{\mathrm{r}}}[\bm{\Phi}_{(i)}]_{m,r}\right),\forall m\in\mathcal{N}$
\STATE~~~ $\nabla_{\bm{\rho}_m}^{(dual)}\gets\nabla_{\bm{\rho}_m}^{(dual)}+\sum_{i=1}^{b}\chi_c^{=}\left(\sum_{t\in\mathcal{N}_{\mathrm{t}}}[\bm{\Psi}_{(i)}]_{m,t},\sum_{r\in\mathcal{N}_{\mathrm{r}}}[\bm{\Phi_i}]_{m,r}\right),\forall m\in\mathcal{N}$
\STATE \textbf{end for}
\STATE $\bm{\lambda}^{(e+1)} \gets \bm{\lambda}^{(e)}+\varepsilon_{\lambda}\nabla_{\bm{\lambda}}^{(dual)}$, $\bm{\mu}^{(e+1)} \gets \bm{\mu}^{(e)}+\varepsilon_{\mu}\nabla_{\bm{\mu}}^{(dual)}$, $\bm{\nu}^{(e+1)} \gets \bm{\nu}^{(e)}+\varepsilon_{\nu}\nabla_{\bm{\nu}}^{(dual)},
\bm{\xi}^{(e+1)} \gets \bm{\xi}^{(e)}+\varepsilon_{\xi}\nabla_{\bm{\xi}}^{(dual)}$, $\bm{\rho}^{(e+1)} \gets \bm{\rho}^{(e)}+\varepsilon_{\rho}\nabla_{\bm{\rho}}^{(dual)}$
\ENDFOR
\end{algorithmic}
\end{algorithm}

\subsection*{C. Summarization of LDLF}
In this subsection, we would like to propose a learning scheme, i.e., LDLF, for training the GBLinks model. Let $\bm{w}$ be the set of trainable parameters of the GBLinks model, which is utilized to generate beam selection policies $\bm{\Phi}$ and $\bm{\Psi}$. The overall training scheme of LDLF is presented in Algorithm 1, where $\mathcal{D}=\{\bm{\kappa}_{(l)}\}_{l=1}^n$ is a set of data with $n$ being the total number of training samples, and $\xi>0$ denotes the step size of optimizer. LDLF is implemented for a fixed number of epochs, and each epoch optimizes the trainable parameters of the GBLinks model using a minibatch of size $b$. After generating the beam selection policies~(line 5), i.e., $\{\bm{\Psi}_{(i)}\}_{i=1}^{b}$ and $\{\bm{\Phi}_{(i)}\}_{i=1}^{b}$, where $\bm{\Psi}_{(i)}$ and $\bm{\Phi}_{(i)}$ are the outputted beam selection policies of the $i$-th sample in each batch. Lagrangian relaxation function~\eqref{Cachenable12} is computed~(line 6) using Lagrangian multipliers $\bm{\lambda}^{(e)},\bm{\mu}^{(e)},\bm{\nu}^{(e)},\bm{\xi}^{(e)},\bm{\rho}^{(e)},$ associated with the current epoch. Then, the optimizer updates the weights of the GBLinks model based on the computed Lagrangian relaxation function~(line 7). In each epoch, we store the subgradient of Lagrangian multipliers with the whole dataset~(lines 8-12). Finally, after each epoch, the Lagrangian multipliers are updated following~\eqref{Cachenable11a}-\eqref{Cachenable11e}~(line 14).

\begin{remark}
When the GBLinks model is trained based on LDLF, the weight parameters of the GBLinks model and Lagrangian multipliers can be adjusted until the constraints of~\eqref{Cachenable08} are satisfied. Specifically, the weight parameters of the GBLinks model is approximated using the Adam optimizer~\cite{Adam2014}. Once the GBLinks model is trained, it can be utilized directly to a new ultra-dense D2D mmWave network instances without considering the constraints and Lagrangian multipliers~\footnote{~Due to the poor interpretability of the GBLinks model, we would like to further study the deep learning model based on deep unfolding to address the considered problem in the future work.}.
\end{remark}

\subsection*{D. Computational Complexity Analysis}
In this subsection, we would like to analyze the computational complexity of the GBLinks model. Given $N$, $N_{\mathrm{t}}, N_{\mathrm{r}}$, and the number of layers of the GBLinks model $K$, the computational complexity of MLP1 is $KN(3\times 2^8N_{\mathrm{t}}N_{\mathrm{r}}+2^9N_{\mathrm{t}}+2^{15}+2^{13})$, the computational complexity of MLP2 is $KN(2^8N_{\mathrm{t}}N_{\mathrm{r}}+2^8N_{\mathrm{t}}+2^{16}+2^{13}+2^6N_{\mathrm{t}})$, the computational complexity of MLP3 is $KN(2^8N_{\mathrm{t}}N_{\mathrm{r}}+2^8N_{\mathrm{t}}+2^{16}+2^{13}+2^6N_{\mathrm{r}})$, and the computational complexity of equation~\eqref{Cachenable13b} is $2^7K(N-1)$. Therefore, the total computational complexity for the GBLinks model is $KN(5\times 2^8N_{\mathrm{t}}N_{\mathrm{r}}+2^{10}N_{\mathrm{t}}+23\times 2^{13}+2^6(N_{\mathrm{t}}+N_{\mathrm{r}}))+2^7K(N-1)$. TABLE~\uppercase\expandafter{\romannumeral3} gives out the comparison of the GBLinks model with other methods from varying $N, N_{\mathrm{t}}$, and $N_{\mathrm{r}}$. It is worth noting that the computational complexity of other methods is analyzed with big-$\mathcal{O}$. Due to the GBLinks model in numerical simulations only utilizes one layer, we set $K=1$. We also assume that the number of the operation times of Step 2 in Algorithm 2 is one. It is not hard to find that GreedyNoSched has the lowest computational complexity, but its performance is not very good, which is shown in Section V. C. Meanwhile, the GBLinks model achieves the lowest computational complexity among the methods except for GreedyNoSched.
\begin{table} [htbp]
  \renewcommand{\tablename}
  \caption{\centering{ TABLE \uppercase\expandafter{\romannumeral3}} \protect \\ \qquad \qquad\qquad COMPUTATIONAL COMPARISON FOR DIFFERENT METHODS}
  \\ \\
  \centering
  \begin{threeparttable}[b]
    \begin{tabular}{| c | c | c | c | c | c |} \hline
 \multirow{2}{*}{$N$}& \multirow{2}{*}{$N_{\mathrm{t}}=N_{\mathrm{r}}$} & \multicolumn{4}{c|}{Methods} \\ \cline{3-6}
  & & GBLinks & SCA-based & Exhaustive Search Scheme & GreedyNoSched  \\
 \hline
 \multirow{2}{*}{$20$} & 16 & $1.07\times 10^7$ & $4.48\times 10^{20}$ & $1.19\times 10^{47}$ & $5.12\times 10^3$  \\
 \cline{2-6}
  & 32 & $3.07\times 10^7$ & $1.14\times 10^{23}$ & $3.17\times 10^{58}$ & $2.05\times 10^4$  \\
 \hline
 \multirow{2}{*}{$30$} & 16 & $1.60\times 10^7$ & $1.14\times 10^{22}$ & $2.19\times 10^{77}$ & $7.68\times 10^3$  \\
 \cline{2-6}
 & 32 & $4.61\times 10^7$ & $2.91\times 10^{24}$ & $6.05\times 10^{88}$ & $3.07\times 10^4$  \\
 \hline
    \end{tabular}
 \end{threeparttable}
\end{table}

\section*{\sc \uppercase\expandafter{\romannumeral5}. Numerical Results}
In this section, we present numerical results to evaluate the effectiveness of the GBLinks model. We first introduce the generation method of the datasets and some parameter settings of LDLF. Then we discuss the convergence and performance of the GBLinks model. Finally, we analyze the generalization ability of the GBLinks model.
\begin{figure}[t]
\renewcommand{\captionfont}{\footnotesize}
\renewcommand*\captionlabeldelim{.}
	\centering
	\captionstyle{flushleft}
	\onelinecaptionstrue
	\includegraphics[width=0.6\columnwidth,keepaspectratio]{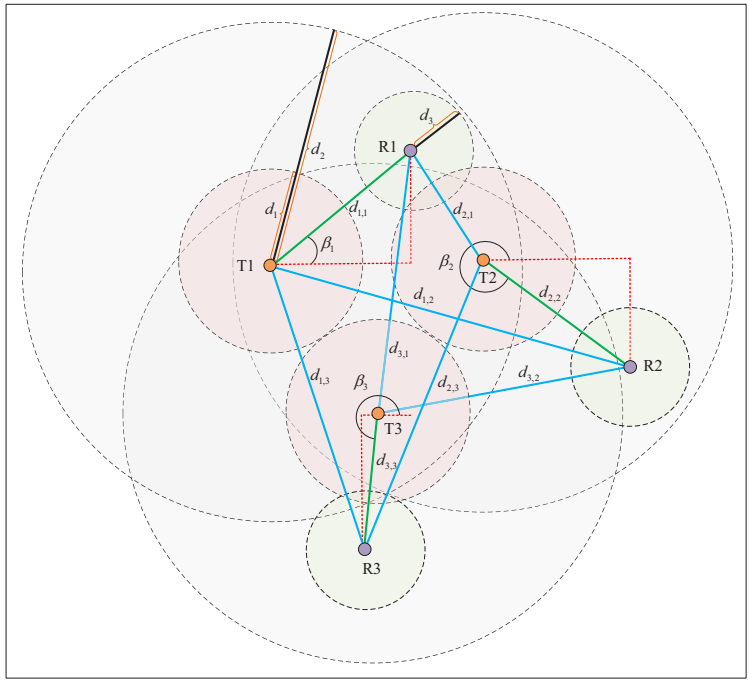}
	\caption{Illustration of generating network topology.}
	\label{networktopology}
\end{figure}

\subsection*{A. Simulation Setup}
The locations of communication pairs are generated randomly. Specifically, the locations of communication pairs are uniformly generated within the region, and the locations of receivers are generated according to a uniform distribution within a pairwise distances of $d_1\sim d_2$ meters from their respective communication pairs. An illustration of generated network topology with three communication pairs is showed in Fig.~\ref{networktopology}. The green line denotes the distance of a communication pair. The red lines denote horizontal and vertical projections of the distance of a communication pair. The blue line denotes the distance of transmitter and receiver associated different communication pairs. The angle $\beta_i\in[0,2\pi], i\in\mathcal{N}$ is also randomly generated following a uniform distribution, which is used to determine the location of receiver corresponding to a transmitter according the generated distance of the communication pair. $d_{m,m}$ and $d_{m,n}$ denote the randomly generated distance of direct link and cross link, respectively.

The predesigned codebook matrices $\mathbf{V}_{\mathrm{r}}$ and $\mathbf{U}_{\mathrm{t}}$ are the Discrete Fourier Transform~(DFT) codebooks~\cite{DFT2017}. We assume that all transmitters have the same transmit power $p_n, n\in\mathcal{N}$, the path amplitudes are assumed to be Rayleigh distributed, i.e., $\alpha_{p,m,n}\sim\mathcal{CN}\left(0,1\right)$, $\rho_{m,n}=d_{m,n}^{-3}$, $m,n\in\mathcal{N}$ and $N_{\mathrm{p}}=2$. The AoDs/AoAs are assumed to take continuous values uniformly distributed in $\left[0, 2\pi\right)$. For easy of notation, we define the SNR as $\mathrm{SNR}=10\mathrm{log}_{10}\left(\frac{P}{\sigma^2}\right)$. For the simplicity, we set $w_m=1, m=1,\cdots,N$. According to the method of generating network topologies, we generate 8000 samples for training and 300 samples for testing, respectively. For the specific parameter setting of the GBLinks model, we set the number of each layer units of MLP1 as $\{3N_{\mathrm{t}}N_{\mathrm{r}}+2N_{\mathrm{t}}, 256, 128, 64\}$, MLP2 as $\{N_{\mathrm{t}}N_{\mathrm{r}}+N_{\mathrm{t}}+128, 256, 128, 64, N_{\mathrm{t}}\}$, MLP3 as $\{N_{\mathrm{t}}N_{\mathrm{r}}+N_{\mathrm{r}}+128,  256, 128, 64, N_{\mathrm{r}}\}$. We consider a batch size of 20 consecutive samples within each drop. The learning rate of the GBLinks model and Lagrangian multipliers are set to $\zeta=1\times 10^{-3}$ and $\varepsilon_{\lambda},\varepsilon_{\mu},\varepsilon_{\nu},\varepsilon_{\xi},\varepsilon_{\rho}=1\times 10^{-6}$, respectively.
\begin{figure}[t]
\renewcommand{\captionfont}{\footnotesize}
\renewcommand*\captionlabeldelim{.}
	\centering
	\captionstyle{flushleft}
	\onelinecaptionstrue
    \includegraphics[width=1.0\columnwidth,keepaspectratio]{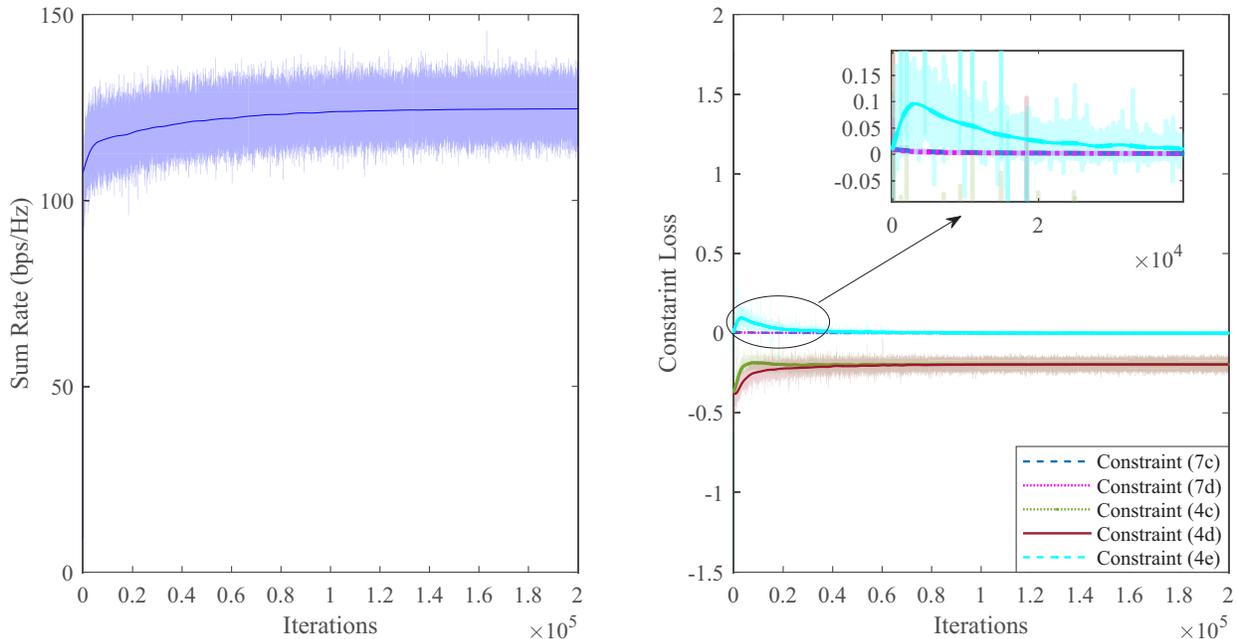}
	\caption{Convergence behavior of the proposed LDLF.}
	\label{LDLF_train}
\end{figure}

\subsection*{B. Effectiveness of LDLF}
In this subsection, we would like to present the effectiveness of training scheme, i.e., LDLF. Specifically, we consider $N=20$ communication pairs with $N_{\mathrm{t}}=16$ and $N_{\mathrm{r}}=16$ in a $50~\text{m}\times50~\text{m}$ region, and $d_1=10~\text{m}, d_2=40~\text{m}$. Fig.~\ref{LDLF_train} illustrates the convergence behavior of LDLF. The left part of Fig.~\ref{LDLF_train} depicts that the GBLinks model trained based on LDLF converges to a stable point with the number of iterations increases, in terms of the weighted sum rate. To examine the feasibility of LDLF, in the right part of Fig.~\ref{LDLF_train}, the constraints~\eqref{Cachenable04c}-\eqref{Cachenable04e} and~\eqref{Cachenable07c}-\eqref{Cachenable07d} are evaluated on average. It can be seen that the GBLinks model becomes feasible after the $1.2\times{10^5}$-th training iterations.

\subsection*{C. Performance Comparison}
In this subsection, we would like to evaluate the performance of the GBLinks model for the case in which the samples in the training dataset and testing dataset have the same number of communication pairs.

\begin{figure}[t]
\renewcommand{\captionfont}{\footnotesize}
\renewcommand*\captionlabeldelim{.}
	\centering
	\captionstyle{flushleft}
	\onelinecaptionstrue
    \includegraphics[width=1.0\columnwidth,keepaspectratio]{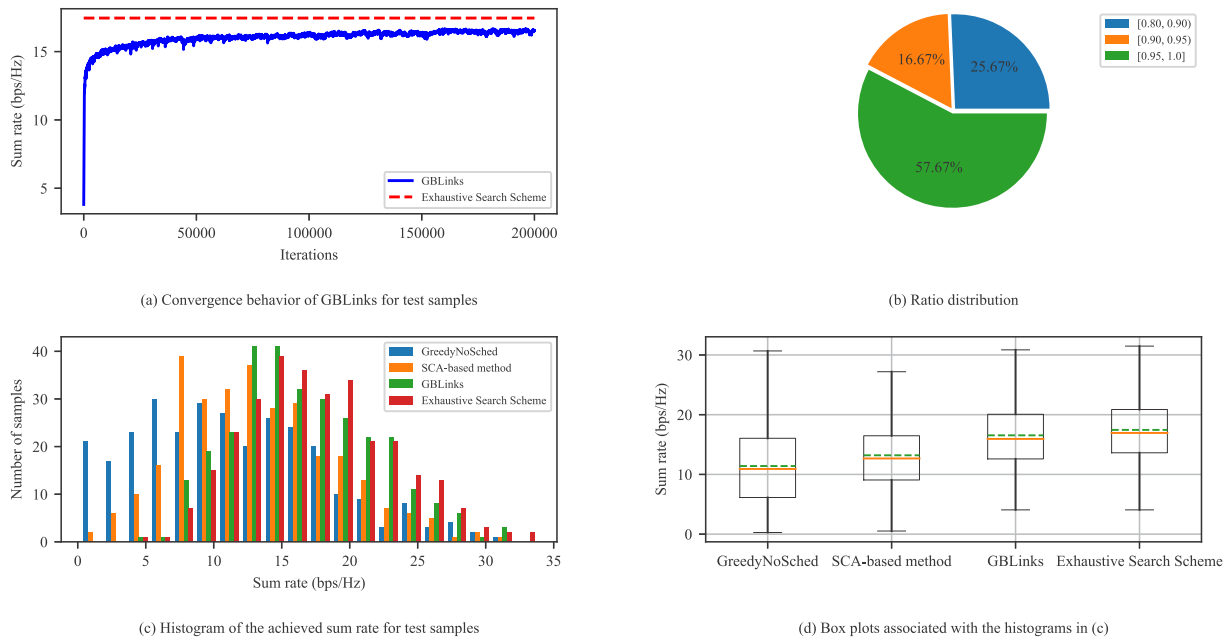}
	\caption{Performance evaluation for the GBLinks model in a smaller D2D mmWave communication network.}
	\label{exhaustVSgblinks_2_16}
\end{figure}

Fig.~\ref{exhaustVSgblinks_2_16} gives out the comparisons between the GBLinks model, the exhaustive search scheme, GreedyNoSched and the SCA-based method for the testing dataset in terms of the weighted sum rate in a smaller-scale D2D mmWave communication network with $N=3$ and $N_{\mathrm{t}}=N_{\mathrm{r}}=8$~\footnote{~Due to the high computational complexity of exhaustive search scheme and the SCA-based method, it is almost undesirable to obtain the optimal beam selection and link activation strategies using exhaustive search scheme or the SCA-based method for large-scale networks. Therefore, we only compare the four methods aforementioned in small-scale wireless network.}. Specifically, the communication pairs are randomly located in a $5~\text{m}\times5~\text{m}$ network coverage region with $d_1=4~\text{m}$ and $d_2=5~\text{m}$. Fig.~\ref{exhaustVSgblinks_2_16}(a) illustrates how the GBLinks model converges to the near-optimal solution as the iteration increases. Then, the percentage pie-chart of ratio $Ra_1$ is illustrated in Fig.~\ref{exhaustVSgblinks_2_16}(b), where $Ra_1=\frac{\text{Weighted sum rate achieved by GBLinks}}{\text{Weighted sum rate achieved by the exhaustive search scheme}}$. We can see that about $74.33\%$ of the testing samples, the value of $Ra_1$ is larger than $0.9$. Fig.~\ref{exhaustVSgblinks_2_16}(c) shows a histogram of weighted sum rates over the entire testing dataset. The empirical distribution of weighted sum rates across different samples shows that the GBLinks model and the exhaustive search scheme obtain a similar weighted sum rate performance for most testing samples. In addition, in Fig.~\ref{exhaustVSgblinks_2_16}(d), we observe that the performance of the GBLinks model is close to that of the exhaustive search scheme in terms of the weighted sum rate value. While the performance of the SCA-based method is a little worse than that of the GBLinks model but is better than that of GreedyNoSched in terms of the weighted sum rate value.

\begin{remark}
As illustrated in Fig.~\ref{exhaustVSgblinks_2_16}, the performance of the GBLinks model is better than the SCA-based method, and can reach about 90\% of the performance of the exhaustive search scheme. And the computational complexity of the GBLinks model is much lower than that of the SCA-based method. Of course, to further improve the performance of the GBLinks model, we can consider the feature design and model structure design of the GBLinks model, which is also our work in the future. In fact, from Fig.~\ref{exhaustVSgblinks_2_16}(a), it is not hard to find that the GBLinks model has converged in $1\times 10^{5}$, which denotes that 250 epoches is sufficient for the training of the GBLinks model. To reduce the number of iterations, we can appropriately increase the value of batchsize when training the GBLinks model according to the performance of the device used. In addition, the GBLinks model is trained in an off-line manner, which does not need to iterate in actual use.
\end{remark}

Table~\uppercase\expandafter{\romannumeral3} lists the performance comparison between the GBLinks model and GreedyNoSched in larger-scale ultra-dense D2D mmWave communication networks in terms of the ratio $Ra_2$, where $Ra_2=\frac{\text{Weighted sum rate achieved by GBLinks}}{\text{Weighted sum rate achieved by GreedyNoSched}}$~\footnote{~Due to memory limitations of existing devices, we do not train the GBLinks model with $N=30$ for $N_{\mathrm{t}}=N_{\mathrm{r}}=32$.}. Specifically, several GBLinks models are separately trained with different training dataset which is generated according to the corresponding configurations of the number of antennas and communication pairs in a $50~\text{m}\times50~\text{m}$ network coverage region with $d_1=10~\text{m}$ and $d_2=40~\text{m}$. For a given network configurations, i.e., the number of antennas and communication pairs, as well as the coverage region, the testing dataset is independently generated with different values of $d_1$ and $d_2$. As shown in Table~\uppercase\expandafter{\romannumeral4}, the GBLinks model achieves a competitive performance advantage compared to GreedyNoSched. It shows that the GBLinks model can be applied in a new scenario with different values of $d_1$ and $d_2$ in a given network configurations. Therefore, the GBLinks model initially has the ability to adapt to the network scenario of mobile communication pairs. For the same antenna configurations and coverage region, the performance gain achieved by the GBLinks model becomes large with an increasing number of communication pairs due to the existing of multiuser diversity gain. However, for the same number of communication pairs and coverage region, the interference mitigation ability of antenna array increases with the increasing number of antennas in the array. Consequence, the performance gain obtained by the GBLinks model becomes weak.

\begin{table} [htbp]
  \renewcommand{\tablename}
  \caption{\centering{ TABLE \uppercase\expandafter{\romannumeral4}} \protect \\  \qquad \qquad PERFORMANCE ANALYSIS OF GBLinks UNDER DIFFERENT NETWORK CONFIGURATIONS}
  \\ \\
  \centering
  \begin{threeparttable}[b]
    \begin{tabular}{| c | c | c | c | c | c | c |} \hline
 \multirow{2}{*}{$\text{SNR~(dB)}$} & \multirow{2}{*}{\makecell[c]{$N$}} & \multirow{2}{*}{$N_{\mathrm{t}}=N_{\mathrm{r}}$} & \multicolumn{4}{c|}{$Ra_2$ with varying $\left(d_1,d_2\right)$} \\ \cline{4-7}
  & & & (5~m, 30~m) & (10~m, 20~m) & (10~m, 40~m) & (20~m, 30~m)  \\
 \hline
 \multirow{3}{*}{$0$} & 20 & 16 & $104.81\%$ & $104.21\%$ & $109.71\%$ & $114.98\%$  \\
 \cline{2-7}
  & 30 & 16 & $113.97\%$ & $115.17\%$ & $124.06\%$ & $128.99\%$  \\
 \cline{2-7}
  & 20 & 32 & 102.99\% & 104.00\% & 106.74\% & 108.76\%  \\
 \hline
 \multirow{3}{*}{$10$} & 20 & 16 & $110.38\%$ & $110.72\%$ & $116.33\%$ & $128.43\%$  \\
 \cline{2-7}
  & 30 & 16 & $114.91\%$ & $116.19\%$ & $125.60\%$ & $131.20\%$  \\
 \cline{2-7}
  & 20 & 32 & 103.05\% & 104.12\% & 106.97\% & 109.18\%  \\
 \hline
    \end{tabular}
 \end{threeparttable}
\end{table}

\begin{figure}[t]
\renewcommand{\captionfont}{\footnotesize}
\renewcommand*\captionlabeldelim{.}
	\centering
	\captionstyle{flushleft}
	\onelinecaptionstrue
    \includegraphics[width=0.8\columnwidth,keepaspectratio]{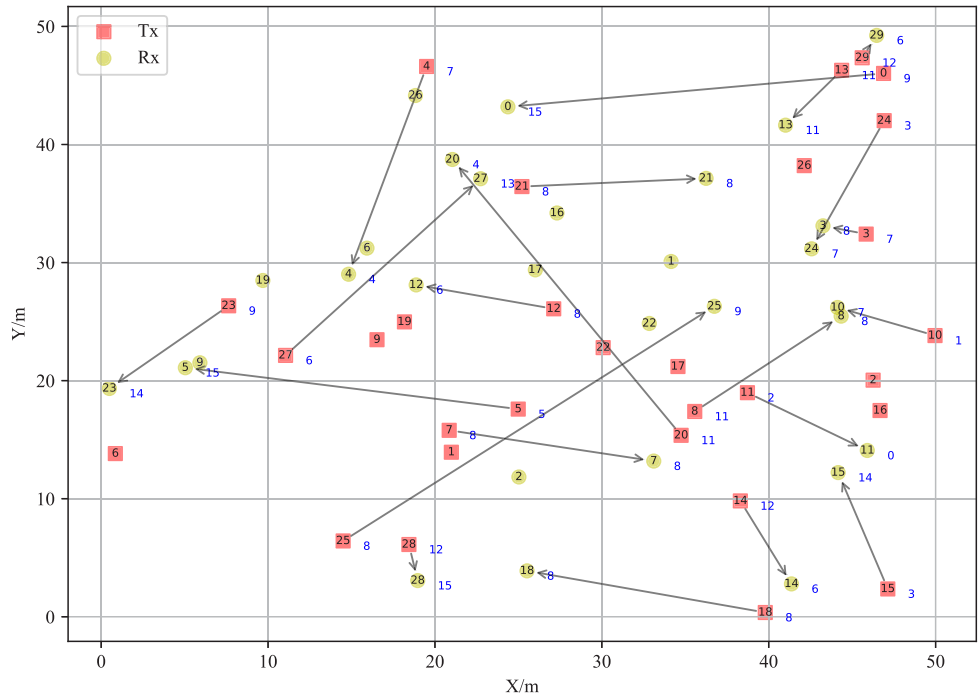}
	\caption{A joint beam selection and link activation consequence.}
	\label{scheduling_output}
\end{figure}

Fig.~\ref{scheduling_output} gives a joint beam selection and link activation consequence with $N=30$ and $N_{\mathrm{t}}=N_{\mathrm{r}}=16$. The numbers within the red square and green circle represent the index of communication pairs. Black arrow lines denote the direct link activated associated with a communication pair, and blue numbers nearing to Rx and Tx indicate the selected beams. From the output of beam selection and link scheduling, $21$ communication pairs of the $30$ communication pairs are activated, and the activated communication pairs have selected the corresponding beams. The ultra-dense D2D mmWave communication network after link activation is relatively sparse, which alleviates the strong interference to a certain extent.

\subsection*{D. Generalization Evaluation of GBLinks}
In this subsection, we further evaluate the generalization ability of the GBLinks model to reveal its ability to adapt to the dynamic network scenarios. Specifically, the GBLinks model is trained separately with different configurations of the number of antennas and communication pairs but with the same network density $N/A=\frac{3}{250}$, $d_1=10~\text{m}$ and $d_2=40~\text{m}$. In the sequel simulations, the value of SNR is set to be $0~\mathrm{dB}$.

\textit{Generalization to varying network densities:}~The testing datasets are generated with different $N$ in a $50~\text{m}\times50~\text{m}$ network coverage region. We compare the GBLinks model with GreedyNoSched in terms of $Ra_2$ and $Ra_3=\frac{\text{Weighted sum rate achieved by GBLinks}}{\text{Weighted sum rate achieved by GBLinks with }N=20}$. As shown in Table~\uppercase\expandafter{\romannumeral5}, for the testing datasets used by the same trained GBLinks model, we observe that the ratio $Ra_2$ is increasing as $N$ increases, i.e., the network density becomes large. The ratio $Ra_3$ does not decrease when the trained model with smaller $N$ is used to evaluate the testing datasets with larger $N$. This implies that the GBLinks model can generalize to the network situations with larger $N$, i.e., more dense D2D mmWave communication network. In addition, for the same testing dataset, the values of the ratios $Ra_2$ and $Ra_3$ obtained by the trained model with smaller $N$ are similar to that obtained by the trained model with larger $N$, respectively. This inspires us to use a GBLinks model with smaller $N$ to deal with the network scenarios with larger $N$ on the premise of ensuring that the network density is close. Therefore, the GBLinks model trained with smaller $N$ has the ability to deal with the more dense D2D mmWave communication networks.

\begin{table}[t]
  \renewcommand{\tablename}
  \caption{\centering{ TABLE \uppercase\expandafter{\romannumeral5}} \protect \\ \qquad \qquad \qquad \qquad GENERALIZATION TEST WITH VARYING DENSITIES}
  \\ \\
  \centering
  \begin{threeparttable}[b]
    \begin{tabular}{| c | c | c | c | c | c | c |} \hline
  \multirow{3}{*}{\makecell[c]{$N$ \\ (Training Model)}} & \multirow{3}{*}{\makecell[c]{$N_{\mathrm{t}}=N_{\mathrm{r}}$}} & \multirow{3}{*}{\makecell[c]{$N$ \\ (Testing Dataset)}} & \multicolumn{4}{c|}{$\left(d_1,d_2\right)$} \\ \cline{4-7}
  & & & \multicolumn{2}{c|}{(10~m, 40~m)} & \multicolumn{2}{c|}{(20~m, 30~m)}  \\
 \cline{4-7}
 & & & $Ra_2$ & $Ra_3$ & $Ra_2$ & $Ra_3$  \\
 \hline
 \multirow{4}{*}{$20$} & \multirow{4}{*}{16} & 20 & $114.61\%$ & \textcolor[rgb]{0.00,0.00,1.00}{100.00\%} & $124.99\%$ & \textcolor[rgb]{0.00,0.00,1.00}{100.00\%} \\
 \cline{3-7}
 & & 30 & $127.05\%$ & \textcolor[rgb]{0.00,0.00,1.00}{119.01\%} & $134.57\%$ & \textcolor[rgb]{0.00,0.00,1.00}{118.12\%} \\
 \cline{3-7}
 & & 40 & $135.04\%$ & \textcolor[rgb]{0.00,0.00,1.00}{136.36\%} & $144.78\%$ & \textcolor[rgb]{0.00,0.00,1.00}{131.16\%} \\
 \cline{3-7}
 & & 50 & $140.79\%$ & \textcolor[rgb]{0.00,0.00,1.00}{150.88\%} & $152.96\%$ & \textcolor[rgb]{0.00,0.00,1.00}{138.36\%} \\
 \cline{1-7}
 \multirow{4}{*}{$30$} & \multirow{4}{*}{16} & 20 & $114.68\%$ & \textcolor[rgb]{0.00,0.00,1.00}{100.00\%} & $121.19\%$ & \textcolor[rgb]{0.00,0.00,1.00}{100.00\%} \\
 \cline{3-7}
 & & 30 & $124.06\%$ & \textcolor[rgb]{0.00,0.00,1.00}{117.61\%} & $128.99\%$ & \textcolor[rgb]{0.00,0.00,1.00}{117.99\%} \\
 \cline{3-7}
 & & 40 & $131.79\%$ & \textcolor[rgb]{0.00,0.00,1.00}{132.69\%} & $139.71\%$ & \textcolor[rgb]{0.00,0.00,1.00}{130.60\%} \\
 \cline{3-7}
 & & 50 & $138.07\%$ & \textcolor[rgb]{0.00,0.00,1.00}{147.20\%} & $149.36\%$ & \textcolor[rgb]{0.00,0.00,1.00}{139.05\%} \\
 \cline{1-7}
 \multirow{4}{*}{$20$} & \multirow{4}{*}{32} & 20 & 107.47\% & \textcolor[rgb]{0.00,0.00,1.00}{100.00\%} & 109.42\% & \textcolor[rgb]{0.00,0.00,1.00}{100.00\%} \\
 \cline{3-7}
 & & 30 & 109.04\% & \textcolor[rgb]{0.00,0.00,1.00}{124.66\%} & 110.27\% & \textcolor[rgb]{0.00,0.00,1.00}{122.25\%} \\
 \cline{3-7}
 & & 40 & 111.24\% & \textcolor[rgb]{0.00,0.00,1.00}{144.68\%} & 111.82\% & \textcolor[rgb]{0.00,0.00,1.00}{139.66\%}\\
 \cline{3-7}
 & & 50 & 111.46\% & \textcolor[rgb]{0.00,0.00,1.00}{166.84\%} & 112.59\% & \textcolor[rgb]{0.00,0.00,1.00}{154.32\%}\\
 \hline
    \end{tabular}
 \end{threeparttable}
\end{table}

\begin{table} [t]
  \renewcommand{\tablename}
  \caption{\centering{ TABLE \uppercase\expandafter{\romannumeral6}} \protect \\ \qquad \qquad \qquad \qquad GENERALIZATION TEST WITH VARYING NETWORK COVERAGE REGION}
  \\ \\
  \centering
  \begin{threeparttable}
    \begin{tabular}{| c | c | c | c | c | c |} \hline
  \multirow{2}{*}{\makecell[c]{$N$ \\(Training Model)}} & \multirow{2}{*}{\makecell[c]{$N_{\mathrm{t}}=N_{\mathrm{r}}$}} & \multirow{2}{*}{Size~($\text{m}^2$)} &  \multirow{2}{*}{\makecell[c]{$N$ \\(Testing Dataset)}} & \multicolumn{2}{c|}{$Ra_2$ with varying $\left(d_1,d_2\right)$} \\ \cline{5-6}
  & & & & (10~m, 40~m) & (20~m, 30~m)  \\
 \hline
 \multirow{4}{*}{$20$} & \multirow{4}{*}{16} & $41~\text{m}\times41~\text{m}$ & 20 & $122.93\%$ & $130.73\%$ \\
 \cline{3-6}
 & & $50~\text{m}\times50~\text{m}$ & 30 & $127.05\%$ & $134.57\%$ \\
 \cline{3-6}
 & & $58~\text{m}\times58~\text{m}$ & 40 & $131.06\%$ & $137.27\%$ \\
 \cline{3-6}
 & & $65~\text{m}\times65~\text{m}$ & 50 & $129.82\%$ & $131.90\%$ \\
 \hline
 \multirow{4}{*}{$30$} & \multirow{4}{*}{16} & $41~\text{m}\times41~\text{m}$ & 20 & $120.19\%$ & $125.64\%$ \\
 \cline{3-6}
 & & $50~\text{m}\times50~\text{m}$ & 30 & $124.06\%$ & $128.99\%$ \\
 \cline{3-6}
 & & $58~\text{m}\times58~\text{m}$ & 40 & $127.43\%$ & $132.92\%$ \\
 \cline{3-6}
 & & $65~\text{m}\times65~\text{m}$ & 50 & $125.28\%$ & $139.46\%$ \\
 \hline
 \multirow{4}{*}{$20$} & \multirow{4}{*}{32} & $41~\text{m}\times41~\text{m}$ & 20 & $108.09\%$ & $111.02\%$ \\
 \cline{3-6}
 & & $50~\text{m}\times50~\text{m}$ & 30 & $109.04\%$ & $110.27\%$ \\
 \cline{3-6}
 & & $58~\text{m}\times58~\text{m}$ & 40 & $109.27\%$ & $111.53\%$ \\
 \cline{3-6}
 & & $65~\text{m}\times65~\text{m}$ & 50 & $108.03\%$ & $110.13\%$ \\
 \hline
    \end{tabular}
 \end{threeparttable}
\end{table}

\textit{Generalization to varying network coverage regions:}~The testing datasets are generated with different $N$ while fixing the network density as the same as the trained models. As shown in Table~\uppercase\expandafter{\romannumeral6}, for the same testing dataset, the ratio $Ra_2$ obtained by the trained model with smaller $N$ is similar to that obtained by the trained model with larger $N$. It verifies again the network scenarios with larger $N$~(more dense) can be handled by the GBLinks model with smaller $N$~(less dense). In addition, combined with Table IV, for the testing datasets with the same configurations, i.e., the number of antennas and communication pairs, but different network coverage regions validated by the same trained model, the performance gain of the testing datasets achieved by the trained GBLinks model in the scenario with smaller network coverage region~(smaller distribution space of communication pairs) is better than that in the scenario with larger network coverage region~(larger distribution space of communication pairs). This is because in a smaller network coverage region, the interference between communication pairs is relatively strong, and the performance of GreedyNoSched becomes worse. In contrast, the effective joint beam selection and link scheduling reduces the interference between communication pairs to a certain extent. In a word, the GBLinks model achieves a competitive performance advantage compared to GreedyNoSched in terms of the ratios $Ra_2$. It shows that the performance of the GBLinks model is stable as the network coverage region varies. Therefore, the GBLinks model has a certain ability to generalize to varying network coverage regions.

\begin{remark}
Based on the simulation results, the input and output of the GBLinks model are only restricted by the number of antennas equipped at the communication pairs. That is, if the number of antennas in the wireless communication system changes, we must retrain the GBLinks model. In contrast, if the number of antennas does not change, from the results of numerical results, we have the following conclusions. On one hand, if the network coverage region does not change, the changes of the network density have little effect on the performance of the GBLinks model. Therefore, in this case, we do not need retrain the GBLinks model. To save the time and memory costs of training the GBLinks model, it is enough to train a GBLinks model based on fewer number of communication pairs in advance. On the other hand, simulation results show that the performance in terms of the weighted sum rate of the GBLinks model, which is trained in a fixed network coverage region, decreases when testing the GBLinks model in a much larger or smaller network coverage region. Hence, when the network coverage region becomes much larger or smaller than that of the current pre-trained GBLinks model, we recommend retraining the GBLinks model according to the current network coverage region.
\end{remark}

\section*{\sc \uppercase\expandafter{\romannumeral6}. Conclusions}
In this paper, we formulated the joint beam selection and link activation problem in ultra-dense D2D mmWave communication networks as a constrained binary integer non-convex optimization problem. In this optimization problem, we just need to optimize the beam indicator variables of transmitter and receiver to finish the joint beam selection and link activation. To address the non-convex and NP-hard problem, we proposed an end-to-end GNN-based learning model named GBLinks, to learn the beam indicator variables, which is trained based on the LDLF proposed. Numerical results show that the proposed GBLinks model can converges to a stable point with the number of iterations increases, in terms of the weighted sum rate. Meanwhile, the GBLinks model can reach near-optimal solution through comparing with exhaustive search scheme in small-scale ultra-dense D2D mmWave communication networks and outperforms GreedyNoSched and the SCA-based method. It also shows that the GBLinks model can generalize to varying network densities and network coverage regions of ultra-dense D2D mmWave communication networks. For the future directions, it is interesting to solve the considered problem based on deep unfolding.

\section*{Appendix}

In this, we propose a SCA-based efficient method to solve problem~\eqref{Cachenable08}. We first formulate the partial Lagrangian dual function of problem as~\eqref{Cachenable15} by introducing Lagrangian multipliers $\vartheta,\delta\geq0$.
\begin{subequations}\label{Cachenable15}
\begin{align}
&\max\limits_{\vartheta,\delta}\min\limits_{\bm{\Psi},\bm{\Phi}}-\sum_{m\in\mathcal{N}}\sum\limits_{r\in\mathcal{N}_{\mathrm{r}}}\sum\limits_{t\in\mathcal{N}_{\mathrm{t}}}w_mR_{m,r,t}+g_1(\varphi)-h_1(\varphi)+g_2(\phi)-h_2(\phi), \label{Cachenable15a} \\
\mathrm{s.t.}~&\textrm{(4c)}-\textrm{(4e)}, \textrm{(7a)}-\textrm{(7b)}, \label{Cachenable15b}
\end{align}
\end{subequations}
where $g_1(\varphi),h_1(\varphi),g_2(\varphi)$, and $h_2(\varphi)$ are defined respectively as~\cite{Poweralloc2016}
\begin{subequations}\label{Cachenable16}
\begin{align}
&g_1(\varphi)\triangleq\vartheta\sum\limits_{n\in\mathcal{N}}\sum\limits_{t\in\mathcal{N}_{\mathrm{t}}}\varphi_{n,t}+\vartheta\left(\sum\limits_{n\in\mathcal{N}}\sum\limits_{t\in\mathcal{N}_{\mathrm{t}}}\varphi_{n,t}\right)^2,  \label{Cachenable16a} \\
&h_1(\varphi)\triangleq\vartheta\sum\limits_{n\in\mathcal{N}}\sum\limits_{t\in\mathcal{N}_{\mathrm{t}}}\varphi_{n,t}^2+\vartheta\left(\sum\limits_{n\in\mathcal{N}}\sum\limits_{t\in\mathcal{N}_{\mathrm{t}}}\varphi_{n,t}\right)^2,  \label{Cachenable16b} \\
&g_2(\phi)\triangleq\delta\sum\limits_{m\in\mathcal{N}}\sum\limits_{r\in\mathcal{N}_{\mathrm{r}}}\phi_{m,r}+\delta\left(\sum\limits_{m\in\mathcal{N}}\sum\limits_{r\in\mathcal{N}_{\mathrm{r}}}\varphi_{m,r}\right)^2,  \label{Cachenable16c} \\
&h_2(\phi)\triangleq\delta\sum\limits_{m\in\mathcal{N}}\sum\limits_{r\in\mathcal{N}_{\mathrm{r}}}\phi_{m,r}^2+\delta\left(\sum\limits_{m\in\mathcal{N}}\sum\limits_{r\in\mathcal{N}_{\mathrm{r}}}\varphi_{m,r}\right)^2.  \label{Cachenable16d}
\end{align}
\end{subequations}
The update of Lagrangian multipliers $\vartheta$ and $\delta$ can be implemented using the sub-gradient method, i.e., $\vartheta^{(\tau+1)}=\vartheta^{(\tau)}+\epsilon\sum\limits_{n\in\mathcal{N}}\sum\limits_{t\in\mathcal{N}_{\mathrm{t}}}\varphi_{n,t}^{(\tau)}\left(1-\varphi_{n,t}^{(\tau)}\right),\delta^{(\tau+1)}=\delta^{(\tau)}+\epsilon\sum\limits_{m\in\mathcal{N}}\sum\limits_{t\in\mathcal{N}_{\mathrm{r}}}\phi_{m,r}^{(\tau)}\left(1-\phi_{m,r}^{(\tau)}\right)$ with $\epsilon\geq0$ being the update step-size, and $\tau$ denoting the $\tau$-th iteration. In the sequel, we focus on the inner optimization of problem~\eqref{Cachenable15} with fixed $\vartheta$ and $\delta$, i.e.,
\begin{subequations}\label{Cachenable17}
\begin{align}
&\min\limits_{\bm{\Psi},\bm{\Phi}}-\sum_{m\in\mathcal{N}}\sum\limits_{r\in\mathcal{N}_{\mathrm{r}}}\sum\limits_{t\in\mathcal{N}_{\mathrm{t}}}w_mR_{m,r,t}+g_1(\varphi)-h_1(\varphi)+g_2(\phi)-h_2(\phi), \label{Cachenable17a} \\
\mathrm{s.t.}~&\textrm{(4c)}-\textrm{(4e)}, \textrm{(7a)}-\textrm{(7b)}, \label{Cachenable17b}
\end{align}
\end{subequations}
It's not difficult to find that problem~\eqref{Cachenable17} is non-convex due to the non-convex objective function~\eqref{Cachenable17a}~\cite{ConvexOpt}. $R_{m,r,t}$ can be equivalently reformulated as
\begin{equation}\label{Cachenable18}
\begin{aligned}
R_{m,r,t}&=f_{m,r,t}(\bar{w})-q_{m,r,t}(\bar{w}), \forall m\in\mathcal{N}, r\in\mathcal{N}_{\mathrm{r}}, t\in\mathcal{N}_{\mathrm{t}},
\end{aligned}
\end{equation}
where
\begin{subequations}\label{Cachenable19}
\begin{align}
&f_{m,r,t}(\bar{w})=\log_{2}\left(\sum\limits_{n\in\mathcal{N}\setminus\left\{m\right\}}\sum\limits_{l\in\mathcal{N}_{\mathrm{t}}}\bar{w}_{m,n,r,l}p_{n}\varrho\left(m,r,n,l\right)+\sigma_{m}^{2}+\bar{w}_{m,m,r,t}p_{m}\varrho\left(m,r,m,t\right)\right), \label{Cachenable19a} \\
&q_{m,r,t}(\bar{w})=\log_{2}\left(\sum\limits_{n\in\mathcal{N}\setminus\left\{m\right\}}\sum\limits_{l\in\mathcal{N}_{\mathrm{t}}}\bar{w}_{m,n,r,l}p_{n}\varrho\left(m,r,n,l\right)+\sigma_{m}^{2}\right). \label{Cachenable19b}
\end{align}
\end{subequations}
where $\bar{w}_{m,n,r,l}=\phi_{m,r}\varphi_{n,l}, \forall m,n\in\mathcal{N},r\in\mathcal{N_{\mathrm{r}}},l\in\mathcal{N}_{\mathrm{t}}$. Then, we can further reformulate problem~\eqref{Cachenable17} as~\eqref{Cachenable20}.
\begin{subequations}\label{Cachenable20}
\begin{align}
&\min-\sum_{m\in\mathcal{N}}\sum\limits_{r\in\mathcal{N}_{\mathrm{r}}}\sum\limits_{t\in\mathcal{N}_{\mathrm{t}}}w_m\left(f_{m,r,t}(\bar{w})-q_{m,r,t}(\bar{w})\right)+g_1(\varphi)-h_1(\varphi)+g_2(\phi)-h_2(\phi), \label{Cachenable20a} \\
\mathrm{s.t.}~&\textrm{(4c)}-\textrm{(4e)}, \textrm{(7a)}-\textrm{(7b)}, \label{Cachenable20b} \\
&\bar{w}_{m,n,r,l}=\phi_{m,r}\varphi_{n,l}, \forall m,n\in\mathcal{N}, r\in\mathcal{N}_{\mathrm{r}}, l\in\mathcal{N}_{\mathrm{t}}, \label{Cachenable20c}
\end{align}
\end{subequations}
In problem~\eqref{Cachenable20}, the optimization variables are $\varphi_{n,l}$, $\phi_{m,r}$, and $\bar{w}_{m,n,r,l}$, $\forall m,n\in\mathcal{N}, r\in\mathcal{N}_{\mathrm{r}}, l\in\mathcal{N}_{\mathrm{t}}$. Exploiting the convexity of $h_1(\varphi)$ and $h_2(\phi)$, the concavity of $q_{m,r,t}$, and first-order Taylor series expansion, we can obtain their low~(upper) boundary approximations as follows
\begin{subequations}\label{Cachenable21}
\begin{align}
&h_1(\varphi)\geq\bar{h_1}(\varphi)\triangleq h_1(\varphi^{(\tau)})+\sum\limits_{n\in\mathcal{N}}\sum\limits_{t\in\mathcal{N}_{\mathrm{t}}}2\vartheta\left(\varphi_{n,t}^{(\tau)}+\sum\limits_{k\in\mathcal{N}}\sum\limits_{l\in\mathcal{N}_{\mathrm{t}}}\varphi_{k,l}^{(\tau)}\right)\left(\varphi_{n,t}-\varphi_{n,t}^{(\tau)}\right), \label{Cachenable21a} \\
&h_2(\phi)\geq\bar{h_2}(\phi)\triangleq h_2(\phi^{(\tau)})+\sum\limits_{m\in\mathcal{N}}\sum\limits_{r\in\mathcal{N}_{\mathrm{r}}}2\delta\left(\phi_{m,r}^{(\tau)}+\sum\limits_{k\in\mathcal{N}}\sum\limits_{l\in\mathcal{N}_{\mathrm{r}}}\phi_{k,l}^{(\tau)}\right)\left(\phi_{m,r}-\phi_{m,r}^{(\tau)}\right), \label{Cachenable21b} \\
&q_{m,r,t}(\bar{w})\leq \bar{q}(\bar{w})\triangleq q_{m,r,t}(\bar{w}^{(\tau)})+\frac{\sum\limits_{n\in\mathcal{N}\setminus\left\{m\right\}}\sum\limits_{l\in\mathcal{N}_{\mathrm{t}}}p_n\varrho(m,r,n,l)\left(\bar{w}_{m,r,n,l}-\bar{w}_{m,r,n,l}^{(\tau)}\right)}{\mathrm{ln}2\left(\sum\limits_{n\in\mathcal{N}\setminus\left\{m\right\}}\sum\limits_{l\in\mathcal{N}_{\mathrm{t}}}\bar{w}_{m,n,r,l}^{(\tau)}p_{n}\varrho\left(m,r,n,l\right)+\sigma_{m}^{2}\right)}.\label{Cachenable21c}
\end{align}
\end{subequations}
In~\eqref{Cachenable21}, $\varphi_{n,t}^{(\tau)}$, $\phi_{m,r}^{(\tau)}$ and $\bar{w}_{m,n,r,l}^{(\tau)}$ represent the solutions obtained at the $\tau$-th iteration for variables $\varphi_{n,t}$, $\phi_{m,r}$ and $\bar{w}_{m,n,r,l}$, respectively. Note that the bilinear form constraints in~\eqref{Cachenable20c} are non-convex, we can solve this challenge via McCormick envelopes~\cite{McCormick2015, Beamforming2021}. Thus, problem~\eqref{Cachenable20} can be reformulated as
\begin{subequations}\label{Cachenable22}
\begin{align}
&\min-\sum_{m\in\mathcal{N}}\sum\limits_{r\in\mathcal{N}_{\mathrm{r}}}\sum\limits_{t\in\mathcal{N}_{\mathrm{t}}}w_m\left(f_{m,r,t}(\bar{w})-\bar{q}_{m,r,t}(\bar{w})\right)+g_1(\varphi)-\bar{h}_1(\varphi)+g_2(\phi)-\bar{h}_2(\phi), \label{Cachenable22a} \\
\mathrm{s.t.}~&\textrm{(4c)}-\textrm{(4e)}, \textrm{(7a)}-\textrm{(7b)}, \label{Cachenable22b} \\
&\bar{w}_{m,n,r,l}\geq0, \forall m,n\in\mathcal{N}, r\in\mathcal{N}_{\mathrm{r}}, l\in\mathcal{N}_{\mathrm{t}}, \label{Cachenable22c} \\
&\bar{w}_{m,n,r,l}\geq\varphi_{n,l}+\phi_{m,r}-1, \forall m,n\in\mathcal{N}, r\in\mathcal{N}_{\mathrm{r}}, l\in\mathcal{N}_{\mathrm{t}}, \label{Cachenable22d} \\
&\bar{w}_{m,n,r,l}\leq\varphi_{n,l}, \forall m,n\in\mathcal{N}, r\in\mathcal{N}_{\mathrm{r}}, l\in\mathcal{N}_{\mathrm{t}}, \label{Cachenable22e} \\
&\bar{w}_{m,n,r,l}\leq\phi_{m,r}, \forall m,n\in\mathcal{N}, r\in\mathcal{N}_{\mathrm{r}}, l\in\mathcal{N}_{\mathrm{t}}. \label{Cachenable22f}
\end{align}
\end{subequations}
In problem~\eqref{Cachenable22}, the optimization variables are $\varphi_{n,l}$, $\phi_{m,r}$, and $\bar{w}_{m,n,r,l}$, $\forall m,n\in\mathcal{N}, r\in\mathcal{N}_{\mathrm{r}}, l\in\mathcal{N}_{\mathrm{t}}$. Problem~\eqref{Cachenable22} is convex and can be solved using classical optimization methods. The detailed steps of solving problem~\eqref{Cachenable15} is outlined in Algorithm 2, where $\eta^{(t)}$ and $\varsigma^{(\tau)}$ denote the objective values of problem~\eqref{Cachenable22} at the $t$-th iteration and the $\tau$-th iteration, respectively. We only activate a single communication pair in the initialization of $\varphi, \phi$ and $\bar{w}$, e.g., $\varphi_{1,1}=1, \phi_{1,1}=1$ and $\bar{w}_{1,1,1,1}=1$, while $\varphi_{m,t}=0, \phi_{m,r}=0$ and $\bar{w}_{m,n,r,t}=0$ if $m,n,r,t\neq1$.

\begin{algorithm}[t]\label{SPSH02}
\caption{Solution of problem~\eqref{Cachenable15}}
\begin{algorithmic}[1]
\STATE Let $t=\tau=0$, $w_m=1, m\in\mathcal{N}$, initialize $\vartheta^{(0)}=0, \delta^{(0)}=0, \phi_{m,r}^{(0)}, \varphi_{n,l}^{(0)}$, and $\bar{w}_{m,n,r,l}^{(0)}$, $\forall m,n\in\mathcal{N}, r\in\mathcal{N}_{\mathrm{r}}, l\in\mathcal{N}_{\mathrm{t}}$, such that constraint~\eqref{Cachenable15b} is satisfied. Calculate the objective value $\eta^{(t)}=\varsigma^{(\tau)}$.
\STATE Let $\tau\gets\tau+1$, solve problem~\eqref{Cachenable22} to obtain $\varphi_{n,l}^{(\tau)}$, $\phi_{m,r}^{(\tau)}$, $\bar{w}_{m,n,r,l}^{(\tau)}$, and $\varsigma^{(\tau)}$, $\forall m,n\in\mathcal{N}, r\in\mathcal{N}_{\mathrm{r}}, l\in\mathcal{N}_{\mathrm{t}}$.
\STATE If $|\frac{\varsigma^{(\tau)}-\varsigma^{(\tau-1)}}{\varsigma^{(\tau-1)}}|\leq\varepsilon$, stop iteration. Otherwise, go to Step 2.
\STATE Let $t\gets t+1$, update $\vartheta^{(t)}=\vartheta^{(t-1)}+\epsilon\sum\limits_{n\in\mathcal{N}}\sum\limits_{t\in\mathcal{N}_{\mathrm{t}}}\varphi_{n,t}^{(t)}\left(1-\varphi_{n,t}^{(t)}\right),\delta^{(t)}=\delta^{(t-1)}+\epsilon\sum\limits_{m\in\mathcal{N}}\sum\limits_{t\in\mathcal{N}_{\mathrm{r}}}\phi_{m,r}^{(t)}\left(1-\phi_{m,r}^{(t)}\right)$, and calculate the objective value $\eta^{(t)}$. If $|\frac{\eta^{(t)}-\eta^{(t-1)}}{\eta^{(t-1)}}|\leq\varepsilon$, stop iteration. Otherwise, go to Step 2.
\end{algorithmic}
\end{algorithm}

\begin{remark}
In Algorithm 2, for given Lagrangian multipliers $\vartheta^{(t)}$ and $\delta^{(t)}$, we iteratively solve problem~\eqref{Cachenable22} until the objective value $\varsigma^{(\tau)}$ converges. In addition, the initialization of $\vartheta^{(0)}$ and $\delta^{(0)}$ should be smaller values that helps to find a better solution of~\eqref{Cachenable22}, the update step-size $\epsilon$ should be a larger value,  $\varphi_{n,t}$ and $\phi_{m,r}$ that are not equal 0 or 1 converge to 0 or 1 more quickly.
\end{remark}

The convergence of Algorithm 2 can be guaranteed by the theory of SCA~\cite{SCA2016}. Note that in~\eqref{Cachenable22}, there are totally $\mathfrak{K}\triangleq N^2N_{\mathrm{r}}N_{\mathrm{t}} + NN_{\mathrm{r}} + NN_{\mathrm{t}}$ optimization variables, and $\mathfrak{J}\triangleq4N^2N_{\mathrm{r}}N_{\mathrm{t}}+NN_{\mathrm{r}}+NN_{\mathrm{t}}+3N$ convex linear constraints. Therefore, the computational complexity of problem~\eqref{Cachenable22} is $\mathcal{O}\left(\mathfrak{K}^3\mathfrak{J}\right)$~\cite{JCB2014}. Thus, the computational complexity of Algorithm 2 is $\digamma\mathcal{O}\left(\mathfrak{K}^3\mathfrak{J}\right)$, where $\digamma$ is the number of the operation times of Step 2 in Algorithm 2.

\begin{small}

\end{small}
\end{document}